\documentclass[12pt,preprint]{aastex}

\begin{document}
\title{Probing the Density in the Galactic Center Region: Wind-Blown
  Bubbles and High-Energy Proton Constraints}

\author{Christopher L. Fryer\altaffilmark{1,2}, Siming
  Liu\altaffilmark{3}, Gabriel Rockefeller\altaffilmark{1,2}, Aimee
  Hungerford\altaffilmark{1}, Guillaume Belanger\altaffilmark{4}}

\altaffiltext{1}{Computer and Computational Science Division, LANL,
Los Alamos, NM 87545}
\altaffiltext{2}{Department of Physics, The University of Arizona,
Tucson, AZ 85721} 
\altaffiltext{3}{Theoretical Division, LANL, Los Alamos, NM 87545}
\altaffiltext{4}{Service d'Astrophysique, CEA-Saclay, 75231 Paris,
France}

\begin{abstract}
  
  Recent observations of the Galactic center in high-energy
  $\gamma$-rays (above 0.1~TeV) have opened up new ways to study this
  region, from understanding the emission source of these high-energy
  photons to constraining the environment in which they are formed.
  We present a revised theoretical density model of the inner 5~pc
  surrounding Sgr A* based on the fact that the underlying structure
  of this region is dominated by the winds from the Wolf-Rayet stars
  orbiting Sgr A*.  An ideal probe and application of this density
  structure is this high energy $\gamma$-ray emission.  We assume a
  proton-scattering model for the production of these $\gamma$-rays
  and then determine first whether such a model is consistent with the
  observations and second whether we can use these observations to
  further constrain the density distribution in the Galactic center.

\end{abstract}


\section{Introduction}

The central 5--10~pc region surrounding Sgr A* is filled with dense
molecular clouds and streamers with a range of characteristic sizes and
densities; these include the circumnuclear disk (CND), the Northern
Ridge, the Western and Southern Streamers, and the 20~km~s$^{-1}$
and 50~km~s$^{-1}$ giant molecular clouds (GMC) and the Molecular
Ridge between them (Herrnstein \& Ho 2004, 2005).  Some features have
densities above $10^5-10^6\ {\rm cm^{-3}}$ while the density of other
features can be as low as $10^3\ {\rm cm^{-3}}$.

The CND is a ring of molecular material with a well-defined inner edge
at $1.5$~pc and an outer edge around $3$--$4$~pc from Sgr A*; it has a
thickness of $\sim 0.4$~pc at the inner edge and expands to a
thickness of $\sim 2$~pc at larger radii (Wright et al. 2001, Vollmer
\& Duschl 2001, Christopher et al. 2005).  Observations of molecular
tracers (HCN and HCO$^{+}$) in gas cores in the CND place the typical
densities of these cores at $10^7$--$10^8$~cm$^{-3}$ and the total
estimated mass of the torus at $10^6$~M$_\odot$ (Christopher et al.
2005).  The volume filling factor of dense clumps in the torus is
low---approximately 1\% (Vollmer \& Duschl 2001).  The torus is
evidently orbiting the black hole; observations by Jackson et al.
(1993) show a clear pattern of radial velocities indicating that the
torus is inclined by $60$--$75$ degrees from the line of sight and is
orbiting around Sgr A*.

The CND and the 50~km~s$^{-1}$ GMC both contain OH (1720 MHz) masers,
which suggest the presence of ongoing interactions between these two
molecular features and the supernova remnant Sgr A East (Yusef-Zadeh
et al. 2001).  Sgr A East is a non-thermal radio shell source centered
$\sim 2.5$~pc to the east of Sgr A*.  It is approximately 8~pc long
(aligned roughly with the Galactic plane) and 6~pc wide (perpendicular
to the Galactic plane).  Several studies (including Mezger et al.
1989, Maeda et al. 2001, Herrnstein \& Ho 2005) have established that
Sgr A East is interacting with other components in the Galactic
center, including ionized gas streamers within the central cavity of
the CND (Sgr A West) and the molecular clouds and streamers
surrounding the remnant.  Mezger et al. (1989) detected a dust ring
surrounding Sgr A East and suggested that this material had been swept
up by an explosive event that produced Sgr A East, and both theorists
and observers since then have suggested several different scenarios
for the formation of this structure.

The trend and mantra among theorists are to simplify, which can often
lead to incorrect conclusions about what is truly happening in the
Galactic Center (GC).  For example, Mezger et al. (1989) noted that
the total mass of material in the dust ring around Sgr A East ($6
\times 10^4$~M$_\odot$) would be comparable to the total mass
contained within the volume of Sgr A East if the density there were
$10^4$~cm$^{-3}$.  Forming the observed dust shell and synchrotron
source via a supernova or other explosive event in such a dense medium
would require a very high energy ($\sim 10^{52}$~erg) and considerable
time ($\sim 10,000$~y or more, depending on details of the formation
scenario).  Such high energies in stellar explosions have been
observed (so-called hypernovae - Nomoto et al. 2004), but these events
are rare ($10^{-5}$--$10^{-6}\ {\rm y^{-1}}$) in any conditions; and
there is strong theoretical (e.g.  Heger et al.  2003) and
observational (e.g.  Priddey et al. 2006) evidence showing that such
events are even more rare, possibly non-existent, in stars with solar
(or above) metallicities.  It is safe to conclude that the scenario
proposed by Mezger et al. (1989) is incorrect, but what factors were
missing in their analysis?  The critical missing piece is the presence
and influence of massive, evolved stars in the central cluster
surrounding Sgr A*.  The winds from these stars almost certainly
pre-date the explosion that produced Sgr A East, so the explosion
occurred not inside a molecular cloud or other dense environment, but
inside the wind bubble associated with the central cluster
(Rockefeller et al.  2005; Fryer et al.  2006a).  By including these
effects, not only do the inferred age and energy of the supernova
explosion decrease to less than 2000~y and approximately
$10^{51}$~erg, respectively, but an additional feature in the X-ray
observations can be explained (Rockefeller et al.  2005).  This is
just one example showing why obtaining an accurate description of the
density profile in the GC is critical to understanding this region.

In this paper, we describe the basic physical features of a wind-blown
bubble density structure in the inner 5-10~pc of the GC. If a constant
density profile is the ``level 0'' approximation for this region, our
representation is only one level higher.  However, because we know the
combined mass loss from the stars near Sgr A*, this model has no more
free parameters than the constant density model.  Indeed, we argue
that the only free parameter is the density into which the winds are
blowing.  This density determines both the density profile and the
extent of the wind blown bubble region.  With this basic density
structure determined by very few observational inputs, we can then add
many of the dense molecular cores that are observed in molecular maps
of the CND and the GC region.  The physical assumption we make is that
these clumps remain intact as the massive star winds blow past them.

With our basic model, we study a potential alternative probe of the
density structure in the GC region: high-energy emission from cosmic
rays. TeV $\gamma$-rays from a point source $5^{\prime\prime}\pm
10_{stat}^{\prime\prime}\pm20_{sys}^{\prime\prime}$ from Sgr A* (HESS
J1745-290) have been observed by a host of observatories: CANGAROO
(Tsuchiya et al. 2004), Whipple (Kosack et al. 2004), HESS (Aharonian
et al. 2004; Hinton et al. 2006) and MAGIC (Albert et al. 2006).  Both
the flux and the spectral index vary from these three observations,
with spectral indexes ranging from $\alpha=2.21 \pm 0.09$ (Aharonian
et al. 2004) to $\alpha=4.6 \pm 0.5$ (Tsuchiya et al.  2004) and
fluxes ranging from 5\% to 40\% that of the Crab Nebula (note these
comparisons are not direct, they correspond to different energy
bands).  In this paper, we adopt as our standard the HESS
measurements: an above-165-GeV flux of $(1.82 \pm 0.22) \times 10^{-7}
{\rm ~ m^{-2} ~ s^{-1}}$ with a spectral index of $\alpha=2.29 \pm
0.05_{stat}\pm0.15_{sys}$, which are also confirmed by MAGIC
observations.

A number of models have been proposed to explain the emission with
sources originating from the entire diffuse 10~pc region of the HESS
PSF to Sgr A* itself (see Aharonian \& Neronov 2004 for a review). The
recent discovery of TeV emission from the GC ridge (Aharonian et
al. 2006) favors a cosmic ray origin, and the fact that its spectral
index is identical to that of HESS J1745-290 suggests that they may
share the same cosmic ray source(s). 

Aharonian \& Neronov (2004), through shocks, and Liu et al.(2006),
through stochastic acceleration, have argued that protons can be
accelerated in the strong magnetic fields generated near the central
supermassive black hole at the heart of Sgr A*.  High energy protons
may also be produced in the supernova remnant shocks of Sgr A East.
These protons will diffuse through the turbulent magnetic fields in
the GC region.  Ultimately, these high-energy protons can scatter
against protons in the wind-blown bubble region, producing pions that
then decay into the gamma-rays we observe.  With our range of density
profiles, we study this production mechanism for gamma-rays.

This paper includes two main thrusts.  We first develop a simple (but
more sophisticated than past work) theorist's model for the density
structure in the inner 5~pc around Sgr A*, based on a wind-blown
bubble produced by the Wolf-Rayet stars orbiting Sgr A*.  The
wind-blown bubble must exist at some level in the GC and we describe
the physics behind this bubble in \S\ 2.  We also show the versatility
of such a basic structure by adding the dense structure of the
circumnuclear disk.  With this density profile, we then move to the
second thrust of the paper, an attempt to understand the high energy
emission arising from the region in the center of the Galaxy and to
use these observations to better understand the density profile.  We
assume that the high energy emission is produced by proton-proton
interactions, one of the options suggested by Aharonian \& Neronov
(2005).  In \S\ 3, we discuss the possible sources for these protons,
placing some initial constraints on this source (and the density)
based on varying assumptions about the observations.  Using our model
density profile, we then calculate the proton scattering through this
medium and present 2-dimensional spatial distributions of the
high-energy emission in \S\ 4.  For the most part, the spatial
distribution is not severely constrained by current observations, but
future instruments could place strong constraints on the density
profile.  In \S\ 5, we study the spectra produced by our mechanism and
show how measurements of the spectral index can also constrain the
spatial distribution.  We conclude with a summary of our results and a
discussion of the future potential of such detailed calculations.

\section{The Density Profile in the Galactic Center}
\label{sec:den}

To determine where our protons scatter, and hence the production site
(and energy distribution) of these photons, we must first determine
the density distribution of the region surrounding Sgr A*.
Rockefeller et al. (2004) modeled the density structure of the region
surrounding the GC by starting with the assumption that the 25
mass-losing stars near the central black hole dominate the gas in the
GC region and the surrounding medium is a vacuum.  Their simulations
found that the interaction of these winds could explain the diffuse
X-rays in the inner 1-2~pc region around Sgr A*.  This picture is only
slightly modified by the young supernova remnant ($\lesssim 2000$~y
old) centered only $\sim 2$~pc from Sgr A* (Rockefeller et al. 2005;
Fryer et al. 2006a).

As one moves away from the central black hole, the picture becomes
more complex.  Because it was closest to the GC, Rockefeller et al.
(2004) included a model for the CND in their simulations.  But several
other structures surround the GC region (Herrnstein \& Ho 2004).  In
addition, the assumption that winds are blowing into a vacuum becomes
less and less accurate as one looks beyond the central 1-2 parsecs.
For the calculation by Rockefeller et al. (2004), which focused on the
central 1--2~pc around Sgr A*, these dense structures at larger
distances have little effect.  When calculating proton scattering
(which can occur 5-6~pc from the central black hole), we can not
ignore their presence so easily.  In particular, we must include the
effect of the interaction between the stellar winds and the
surrounding cloud of molecular gas.  For our current calculation, we
will assume the following model for the formation of the GC: the
region around the GC is produced by interaction of the stellar winds
in a large cold molecular cloud region with densities between
$10^2-10^5\ {\rm cm^{-3}}$.  The current structures that have not been
swept out by this wind are denser clumps in the large cloud of gas.
The winds blow a bubble in this cloud.

The existence of this wind-bubble region is without doubt.  Such
structures are seen in all star-forming regions and are required for
our understanding of any situation where massive stars have played a
role in defining the surroundings: from star-forming regions
(e.g. Brown, Hartmann, \& Burton 1995) to supernova remnants
(e.g. Dwarkadas 2005) and to the emission from $\gamma$-ray bursts
(e.g. Chevalier \& Li 1999).  In all of these examples, the wind-blown
bubble defines the density profile of the region.  It will be no
different for the area immediately surrounding Sgr A*, where the
roughly two dozen mass-losing stars drive an incredible total wind of
$3\times10^{-3}~{\rm M_\odot ~ y^{-1}}$ at a mean velocity of over
$700 {\rm ~ km ~ s^{-1}}$, which dominates the gas in the inner
5~pc region.

\begin{figure}
\plotone{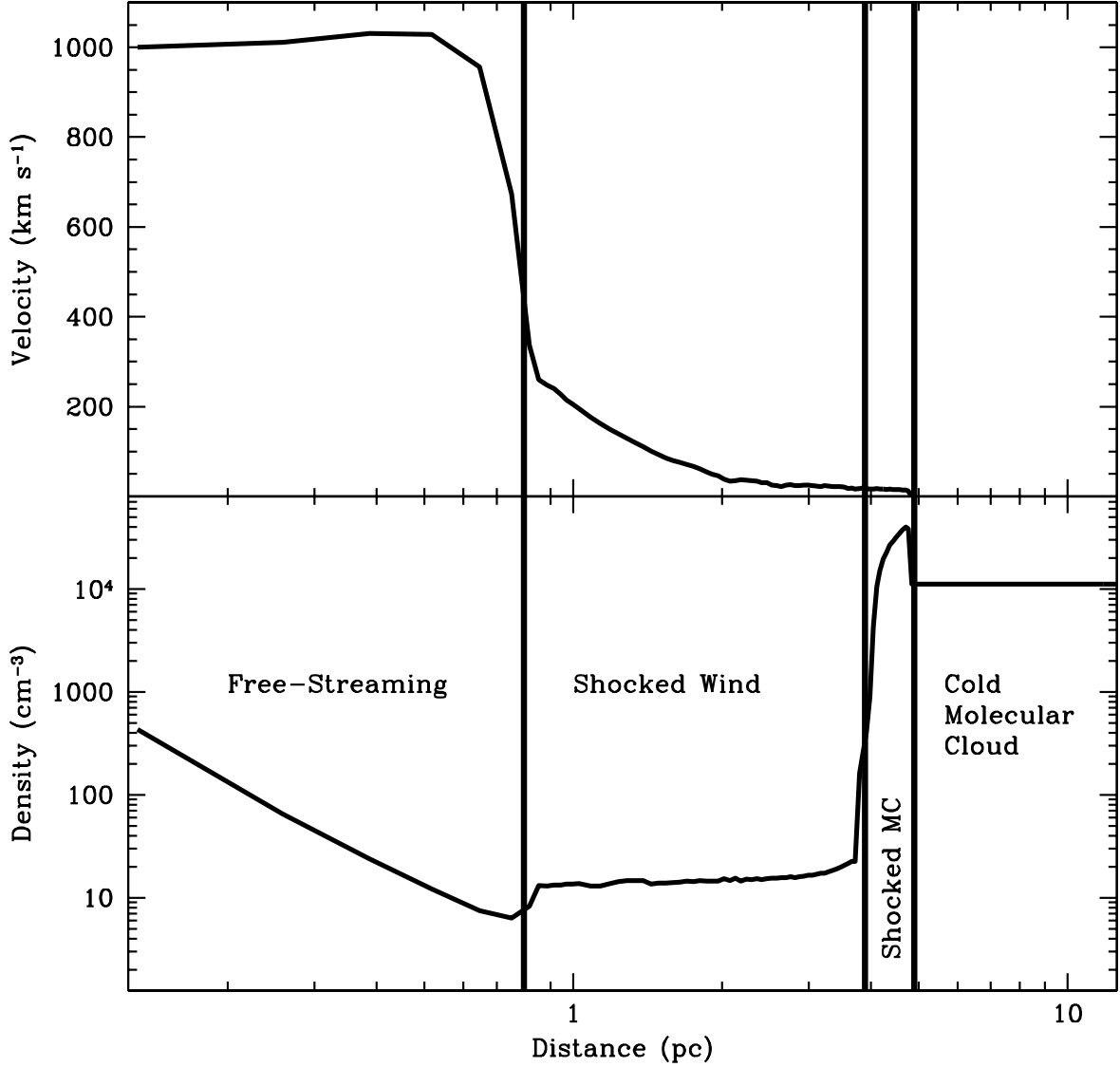}
\caption{Velocity (top) and density (bottom) versus radius of the
wind-blown bubble created by our standard set of conditions: a
$3\times10^{-3} {\rm M_\odot ~ s^{-1}}$, $1000~ {\rm km ~ s^{-1}}$ 
wind blowing into a density of $10^4 {\rm cm^{-3}}$.  This structure 
is 110,000~y after the onset of the wind.  We have separated the 
structure into 4 regions:  free-streaming wind material (with its 
density falling off as the square of the radius), shocked wind 
material, shocked molecular cloud material (the outer edge of this 
region marks the end of the wind bubble), and unshocked (``cold'') 
molecular cloud material.}
\label{fig:bubble}
\end{figure}

For our simplest theorist model, we develop a wind solution for the GC
region.  The characteristics of a wind-blown bubble are determined by
3 parameters: the density into which the wind is blowing and the total
mass loss rate and velocity of the wind.  Fortunately, two of these
parameters are fairly well known: the total mass loss and velocities
of the winds.  In our simple theorist model, we model the mass loss of
all the winds of these stars as a single source of matter.  Far from
the inner 10\arcsec of the GC, such an assumption is easily justified.
The structure of such a simple wind-blown bubble consists of
free-streaming, shocked wind, shocked molecular cloud, and cold
molecular cloud regions (Fig.~\ref{fig:bubble})\footnote{Our cold
molecular cloud is assumed to have sound speeds of roughly
30-50\,km\,s$^{-1}$, so its temperature is not zero.}.  The structure
shown in Fig.~\ref{fig:bubble} was calculated using a 1-dimensional
hydrodynamics code (Fryer et al. 1999; Fryer, Rockefeller \& Young
2006b) assuming a density in the surrounding medium of $n = 10^4\ {\rm
cm^{-3}}$, a mass loss rate of $\dot{M}_{\rm wind}=3\times10^{-3}~{\rm
M_\odot ~ y^{-1}}$ and a wind velocity of $v_{\rm wind}=1000\ {\rm km
s^{-1}}$.

These 4 regions are all well understood under our 3 free parameters
that determine their extent and values. For instance, the extent of
the free-streaming region depends primarily upon the mass loss rate
and the density of the surrounding medium.  For the most part, we will
assume that we know the mass loss rate and wind velocity, and focus
our studies on our most uncertain parameter: the density of the
surrounding medium.  Fryer et al. (2006b) found that the extent of the
free-streaming region ($r_{\rm free-streaming}$) is inversely
proportional to the square root of the density of the surrounding
medium ($\rho_{\rm MC}$): $r_{\rm free-streaming} \propto \rho_{\rm
MC}^{-1/2}$.  The density of all the remaining regions is then
proportional to the density of the surrounding medium (e.g. $\rho_{\rm
shocked wind} \propto \rho_{\rm MC}$).

\begin{figure}
\plotone{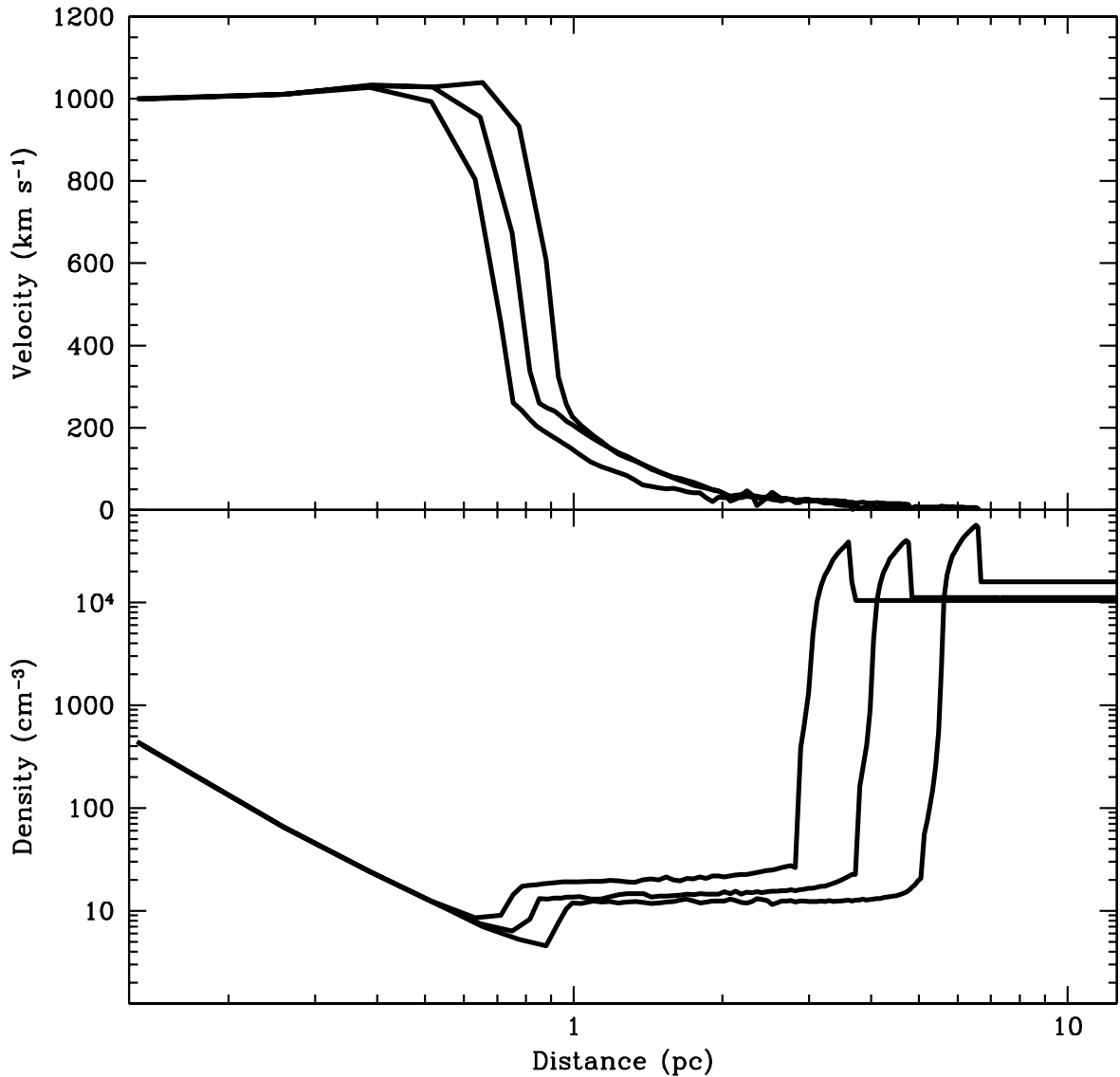}
\caption{Velocity (top) and density (bottom) of our standard set of
wind conditions for 3 different times: 60,000~y, 110,000~y, and
230,000~y.  The edges of all the regions move outward with time.  For
instance, the outer edge of the shocked molecular cloud region (edge
of the wind-blown bubble) moves 3.7~pc in the first 60,000~y,
1.2~pc in the next 50,000~y (a total of 4.9~pc in 110,000~y), and
1.7~pc in the next 120,000~y (a total of 6.6~pc in 230,000~y).  
If we were modeling turbulence and cooling in the shock, when the 
outward motion of the shock decelerated below the turbulent velocity, 
the wind bubble would stall.  Because our molecular cloud is cool (and 
because we include its self-gravity), it gradually compresses, causing a slight 
rise in the density at late times.}
\label{fig:rhovt}
\end{figure}

\begin{figure}
\plotone{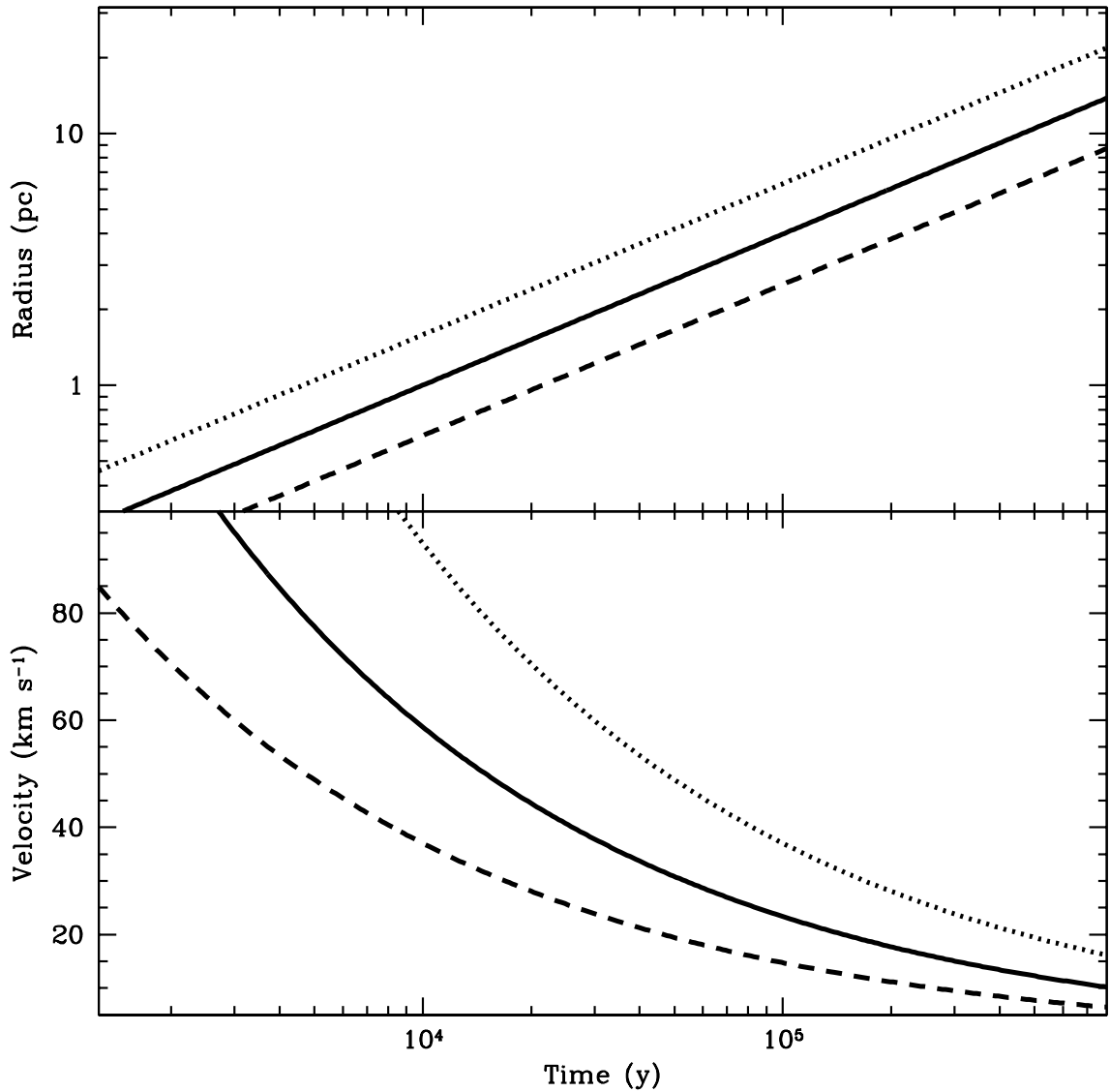}
\caption{Radius and expansion velocity of the wind-blown bubble as a
function of time for three values of $\dot{M}_{wind} v^2_{\rm
wind}/\rho_{\rm MC}$:  our standard values (solid line), a factor 
of 10 increase in this value (dotted line), and a factor of 10 decrease 
in this value (dashed line).  When the velocity of expansion drops below 
the turbulent (or sound speed velocity), the effective expansion of the 
bubble halts.  Note that this happens before the wind-blown bubble is 
100,000~y old.  We can safely assume that the bubble produced by the 
stars surrounding Sgr A* has reached this steady state.}
\label{fig:rvvst}
\end{figure}

Of course, the wind-blown bubble structure will evolve with time (see
Fig.~\ref{fig:rhovt}).  The radius ($r_{\rm bubble}$) and expansion
velocity ($v_{\rm bubble}$) of this bubble can be estimated through
dimensional analysis:
\begin{equation}
\label{eq:rext}
r_{\rm bubble} \propto (\dot{M}_{\rm wind} v^2_{\rm wind}/\rho_{\rm
MC})^{1/5} t^{3/5},
\end{equation}
and 
\begin{equation}
v_{\rm bubble} \propto (\dot{M}_{\rm wind} v^2_{\rm wind}/\rho_{\rm
MC})^{1/5} t^{-2/5},
\end{equation}
where $t$ is the time since the onset of the winds.  We can use our
simulation to calculate the proportionality coefficient.  This factor
is roughly 0.6 for $r_{\rm bubble}$ (eq.~\ref{eq:rext}).  Figure
~\ref{fig:rvvst} shows the evolution of the wind bubble as a function
of time for our standard set of values: $\dot{M}_{\rm wind} =
3\times10^{-3}~{\rm M_\odot ~ y^{-1}}$, $v_{\rm wind} = 1000\ {\rm km
  s^{-1}}$ and $\rho_{\rm MC} = n m_p=10^4 m_p\ {\rm cm^{-3}}$ (solid
line), where $m_p$ is the proton mass, and a simulation where the
value of $\dot{M}_{wind} v^2_{\rm wind}/\rho_{\rm MC}$ is a factor of
10 higher (dotted) and a factor of 10 lower (dashed).  The expansion
of this bubble continues until the outward motion of the bubble slows
to the speed of the turbulent velocities of the surrounding medium in
which it is traveling.  At this point the bubble stalls.

How can we compare this bubble model to the GC?  The size of Sgr A
East is roughly at the edge of the wind-blown bubble.  If we assume
our standard model and a turbulent velocity of $30\ {\rm km ~
  s^{-1}}$, then the extent of the bubble is 2.8~pc.  If we assume the
wind velocity and mass loss is isotropic (not strictly true) and our
only free parameter is the molecular cloud density, a density of
$2.2\times 10^3 ~ {\rm cm^{-3}}$ would be required to produce the 6~pc
edge of Sgr A East.  If we instead assume a $50 {\rm km ~s^{-1}}$
turbulent velocity, the required densities would be $4.8 \times10^2 ~
{\rm cm^{-3}}$ and $1.9\times10^3 ~ {\rm cm^{-3}}$ for the 6~pc and
3~pc edges, respectively.  This establishes the first of many
constraints we can use to help us understand the densities in the GC.

\begin{figure}
\plotone{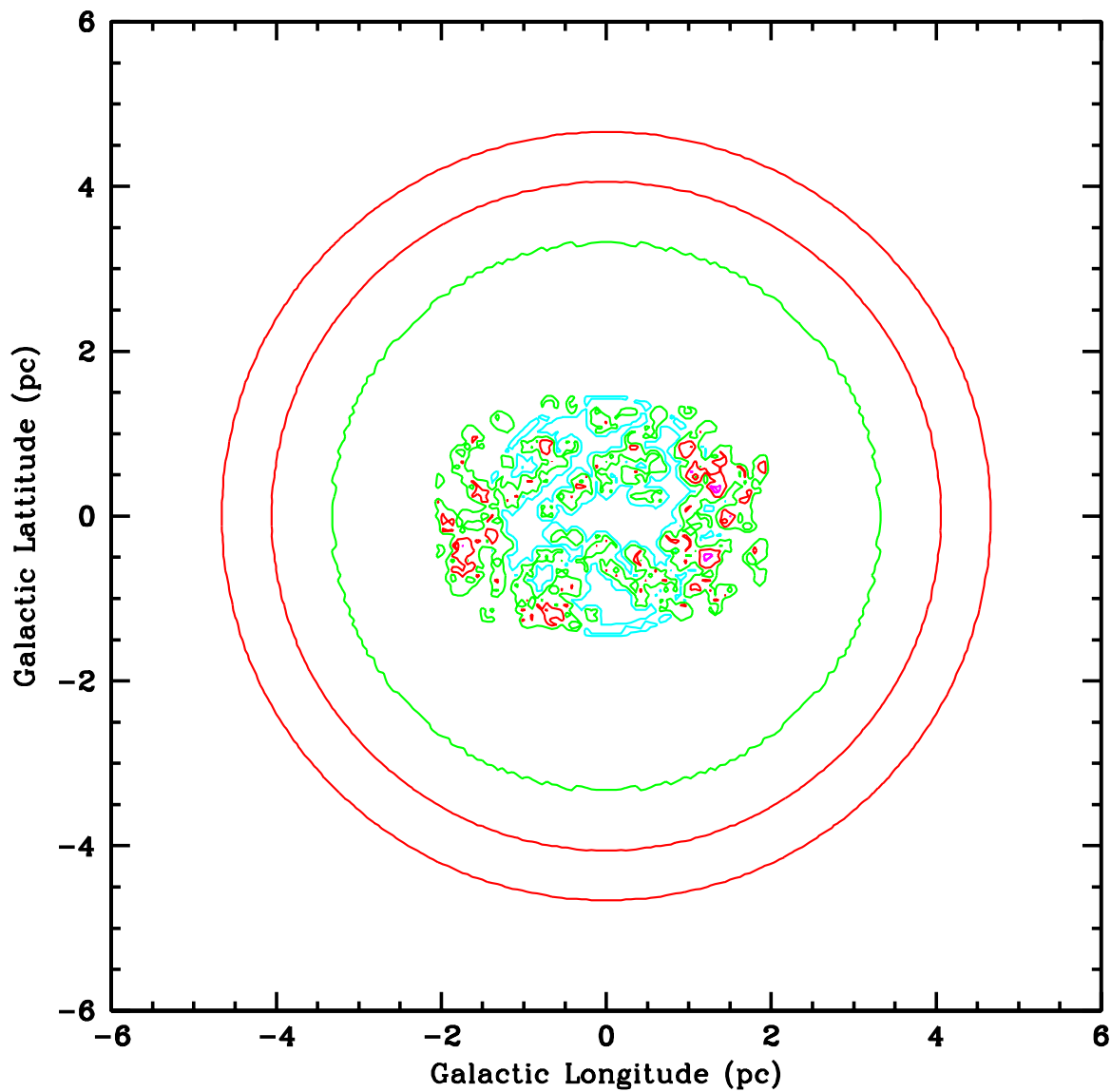}
\caption{Density contours showing the density averaged along the line of sight of 
our 6~pc cubed region surrounding Sgr A* in the projected longitude/latitude
plane.  The contours correspond to the following average densities:
magenta ($>10^4 {\rm cm^{-3}}$), red ($10^3-10^4 {\rm cm^{-3}}$), green
($10^2-10^3 {\rm cm^{-3}}$), and cyan ($<10^2 {\rm cm^{-3}}$).  The
torus causes an increased density in the central 2~pc region.  
The edge of the wind-blown bubble causes a density enhancement in 
a ring near 4~pc.}
\label{fig:density}
\end{figure}

Clumps exist in this wind profile, the largest being the dense cores
in the CND.  Any model of the GC region must also include these
structures.  In this paper, we include only the circumnuclear disk in
modeling the high-energy $\gamma$-ray emission.  A number of other
dense filaments (a.k.a. ``streamers'') and clumps exist in this
central region and we do not suggest that we are presenting a complete
picture of the GC, but instead are showcasing what we can learn by
using the emission from proton scattering to image the GC region.  We
use the circumnuclear disk as an example that the density perturbation
can be imaged by this technique.  The combined density structure of
our wind plus bubble region and the circumnuclear torus is shown in a
2-dimensional image in Figure~\ref{fig:density}.  The dense clumps in
the center are the projected image of the circumnuclear disk.

\section{Proton Propogation:  Effective Proton Opacities}

The scattering of high-energy protons with protons in our density
field ultimately produces the gamma-rays that we observe in detectors
such as HESS.  As we shall see below, the cross-section for this
scattering is quite low and this cross-section alone would produce
very few gamma-rays.  However, the protons can be effectively trapped
by the high magnetic fields in this region and their slow diffusion
out of the GC allows ample time for the protons to
scatter, producing the pions that ultimately emit gamma-rays near Sgr
A*.  Here we discuss our model for the propogation of high-energy
protons through magnetic fields, focusing on the uncertainties in this
model primarily based on the uncertainties in the magnetic field
topology in the GC region.  We then describe the current
understanding of proton-proton scattering and the simplifications we
employ for our calculation of this scattering opacity.

\subsection{Magnetic Field Effects}
\label{sec:turb}

\begin{figure}
\plotone{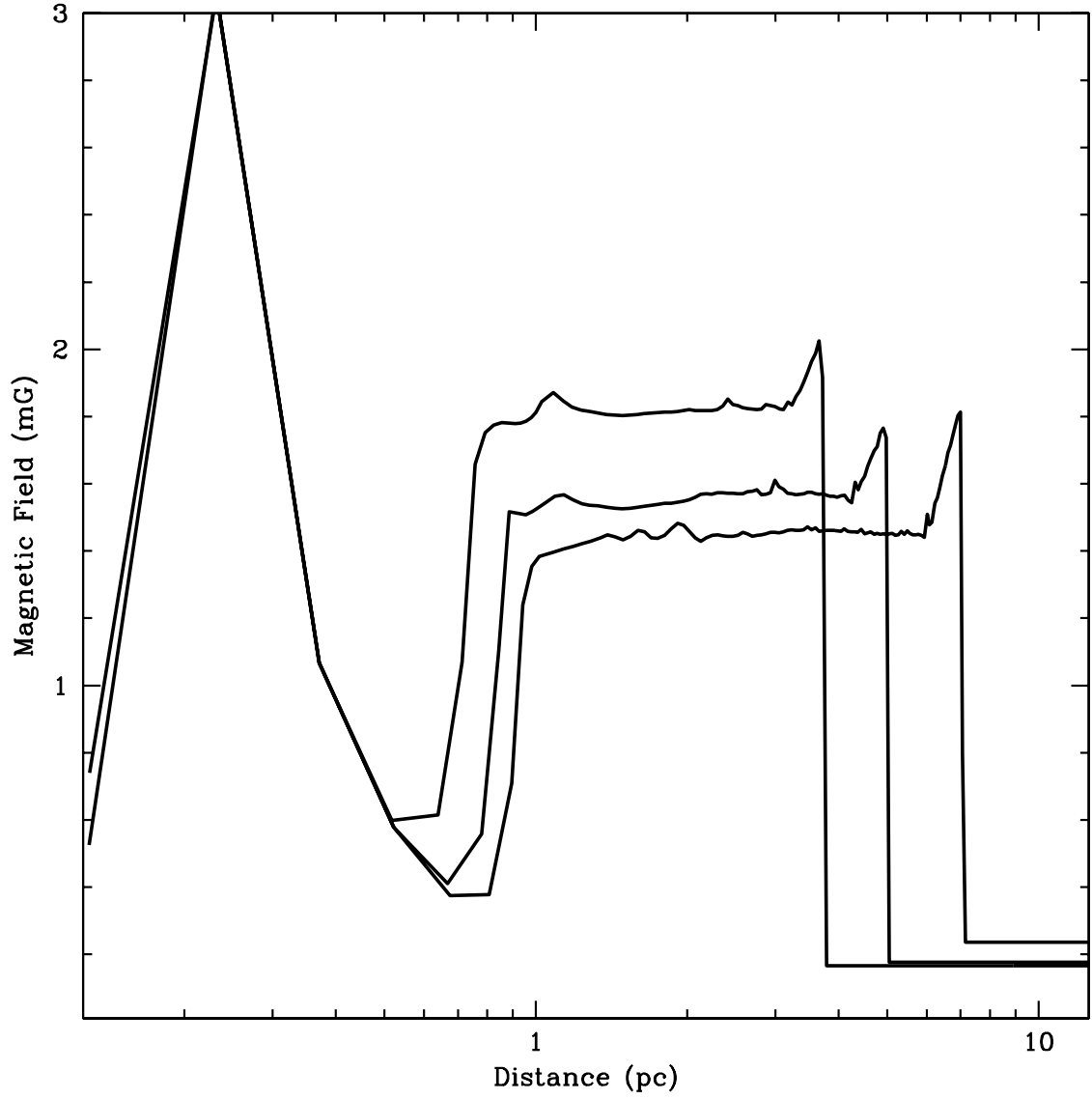}
\caption{Magnetic field in mG for the wind blown bubble assuming the
magnetic field energy is in equipartition with the thermal energy of
the matter.  The shocked regions have high values as their
temperatures, and hence energies, are high.}
\label{fig:mag}
\end{figure}

The propogation of a high-energy proton through a magnetic field can
be calculated if the magnetic field strengths and distributions are
well-defined.  Unfortunately, especially for magnetic fields produced
by turbulence and/or shocks, neither of these are true.  Due to the
dynamic mixing and shocks in this region, magnetic field strengths $B$
in this region are likely to be comparable to those found in typical
dense clouds and, in typical star-forming regions, these magnetic
fields are within an order of magnitude of their equipartition value
(Padoan et al. 2004).  But there is some observational evidence that
the GC fields are much lower than this value (LaRosa et al. 2005).
The magnetic field in our wind-blown region is shown in
Fig.~\ref{fig:mag}.  For the most part, we assume the equipartition
value for the magnetic field strength with coherent lengths below the
size of our spatial resolution.

Assuming equipartition gives us a net magnetic field, but it still
does not provide a distribution for the magnetic field.  Can we assume
that the power of the magnetic field as a function of scale-length of
that field follows a Kolmogorov spectrum?  Although this is commonly
assumed (Blasi \& Colafrancesco 1999), it is not necessarily the
correct answer.  We instead choose to parametrize the magnetic field
energy distribution assuming a power law and vary the power to
determine the dependence of our results on this uncertainty.

We use the derivation of Petrosian and Liu (2004) to parameterize 
the uncertainties in our magnetic field topology.  With this 
derivation, the mean free path of particles in our magnetic 
field has the following form:
\begin{equation}
\lambda_{\rm sc} = c\gamma^{2-q}f(\beta_{\rm A})){8\over \pi (q-1)
f_{\rm turb} c k_{\rm min}}\left({ck_{\rm min}\over
\Omega}\right)^{2-q}~,
\label{eq:tausc}
\end{equation}
where $c$ is the speed of light, $\beta_{\rm A} = v_{\rm A}/c$ is the
Alfven velocity in the plasma in units of the speed of light, $\gamma$
and $\Omega$ are the Lorentz factor and the non-relativistic
gyrofrequency of the particle, respectively.  For our values of the
Alfven velocity, the function $f(\beta_{\rm A})$ is only weakly
dependent upon $\beta_{\rm A}$ (it scales roughly proportional to the
log of $\beta_{\rm A}$ and, for the purposes of this simulation, we
set it equal to a constant: $f(\beta_{\rm A})\approx5$.  $f_{\rm
turb}$ is the energy ratio between the turbulence and the magnetic
field (roughly 1).  $k_{\rm min}$ is the minimum wave number of the
magnetic field (the corresponding maximum scale length of the magnetic
field is $\lambda_{\rm max}=2 \pi/k_{\rm min}$) and $q$ is a power
index corresponding to the distribution of this magnetic field.

This formula boils down our uncertainties in the magnetic field
topology into two parameters: $k_{\rm min}$ (or $\lambda_{\rm max}$)
and $q$.  It implicitly assumes a power law for the magnetic field
energy distribution in the range that determine the propogation of our
high-energy protons.  Typically, we would expect the power law to
break when the gyroradius of the particle exceeds the scale length of
the magnetic fields.  As we increase the proton energy, the gyroradius
increases, but even for a 100~TeV proton, the gyroradius is only
$3.3\times 10^{14}(B/1{\rm mG})^{-1}$~cm and, based on the
calculations by Rockefeller et al. (2004) of the region surrounding
Sgr A*, we expect the scale-length of the magnetic field to be larger
than this value: $\lambda_{\rm max} \gtrsim 10^{16} ~ {\rm cm}$.  So
it is likely that only a single power is needed for our simplified
calculations.

For this paper, we will study two values of $q$: 1.5 and 2\footnote{A
  Kolmogorov spectrum assumes a $q$ value of 5/3 (see, e.g., Blasi \&
  Colafrancesco 1999) and our choices bound this value.}.  If we
assume $q=2$, the scattering length ($\lambda_{\rm sc}$) is
independent of the energy of the proton ($\gamma$) and the magnetic
field strength.  With $\beta_{\rm A} << 1$ and $f_{\rm turb} \approx
1$, equation~\ref{eq:tausc} becomes:
\begin{equation}
\lambda_{\rm sc} \approx 2 \lambda_{\rm max}
\label{eq:tausc-q2}
\end{equation}
Our only other free parameter, then, is the magnetic field scale
length ($\lambda_{\rm max}$), and we test the dependence of the
results on this parameter by running calculations with two values for
$\lambda_{\rm max}$: $10^{15}$ and $10^{17}$~cm.  If we assume
$q=1.5$, we retain a mean-free path dependence both on the average
magnetic field strength $B$ and the proton energy $E_{\rm prot}$:
\begin{equation}
\lambda_{\rm sc} \approx 2\times10^7 (\lambda_{\rm max}/{\rm cm})^{1/2} 
(B/0.1mG)^{-1/2} (E_{\rm prot}/10~TeV)^{1/2} {\rm cm}. 
\label{eq:tausc-q1.5}
\end{equation}
The bulk of our calculations use $q = 1.5$.  We obtain our average
magnetic field strength set to the equipartition value from our
wind-blown bubble calculations and a value of $\lambda_{\rm max}$ of
$10^{15} ~ {\rm cm}$.  In this paper, these mean free paths are
modeled as effective scattering mean-free paths.  The ``scattering''
is assumed to be isotropic and isoenergetic and serves mostly to
redirect the protons, causing them to diffuse out of the Galactic
Center region.

Our assumption that the magnetic field energy density is equal to the
thermal density of our gas implicitly assumes that the largest
scale-length of our magnetic fields is below the resolution of our
models.  But in much of the Galaxy, the magnetic fields are generated
by the Galactic disk, not by small-scale turbulence, and the scale
length of these fields can be large (parsec scales or above).  In the
GC region we consider for much of this paper, we focus on the
magnetic-fields generated in our shocked wind bubble or in dense
molecular clouds and it is likely that in this region, the scale
length of the magnetic fields will be short compared to the size-scale 
of the hydrodynamics simulations (Rockefeller et al. 2004) 
that set up our initial conditions.

\subsection{Proton Scattering}

Magnetic fields may determine the propagation of the high-energy
protons, but it is the proton-proton scattering cross-section that
produces the pions that decay into the high-energy photons we observe.  
Suzuki et al. (2005) have fit the observed opacities for proton 
scattering with a relatively simple empirical function:
\begin{equation}
\sigma_{\rm pp}(E_{\rm prot}) = S_0 (c/v_{\rm proton})^\kappa E_{\rm prot}^\delta 
(1-e^{-\mu E_{\rm prot}}),
\label{eq:pp}
\end{equation}
where 
\begin{equation}
\delta = 4.00 \times 10^{-2} + 5.38 \times 10^{-4} \times E_{\rm prot}^{0.240},
\end{equation}
$S_0 = 2.6 \times 10^{-26} ~ {\rm cm^{-2}}$, $\kappa=2.23$, $\mu =
1.07$ and $E_{\rm prot}$ is the proton energy in GeV and $v_0$ is the
proton's velocity.  Most of the experimental data constraining this 
cross-section is below 100~GeV, but a few data points out to 2~TeV 
exist and this empirical formula fits this data points within the error 
bars.  Above 2~TeV, the opacity is extremely uncertain.

Due to these uncertainties, for most of the calculations in this
paper, we assume a constant value for this opacity ($3.2 \times
10^{-26} ~ {\rm cm^{2}}$), leading to a mean free path for
proton-proton scattering of:
\begin{equation}
\lambda_{\rm mean free path}^{\rm pion} = 3.125 \times 10^{25} n^{-1}_{\rm H} cm^{-2}.
\label{eq:ppflat}
\end{equation}
We do include 1 calculation using the full description from
equation~\ref{eq:pp}, but it is a small effect compared to our other
uncertainties.  We will discuss the effect of such an assumption
in our conclusions.  Despite the fact that the proton is essentially
performing a random walk away from the central black hole due to
magnetic fields, most of our high energy protons do not actually
scatter and decay until they hit the shocked interstellar medium
region of the wind-blown bubble or the dense circumnuclear torus.  For
the bulk of our calculations, we will assume that the proton loses so
much energy after this pion production that it is no longer able to
produce further pions.  Pions decay rapidly, producing high energy
photons.

\section{Simulating Pion Production}
\label{sec:protons}

Since the discovery of the point source HESS J1745-290 in the GC,
several models have been proposed.  The $\gamma$-rays can be produced
either via inverse Compton scattering by relativistic leptons (Atoyan
\& Dermer 2004; Quataert \& Loeb 2005; Wang et al.  2006) or via
hadronic processes (Aharonian \& Neronov 2005; Liu et al.  2006).  The
recent detection of diffuse $\gamma$-ray emission in correlation with
the Galactic molecular cloud density distribution suggests that there
is a cosmic ray source in the GC and therefore favors the latter
scenario (Aharonian et al. 2006).

Two features of this emission are worth special attention: its
spectral index is identical to that of HESS J1745-290, and its
brightness is more than two orders of magnitude lower than that of
HESS J1745-290 (Aharonian et al. 2004, 2006). Given the large source
size of the diffuse emission, its total flux is actually about a
factor of 2 higher. These results set strict constraints on the
possible cosmic ray sources in the Galactic Center: Sgr A*, pulsar
wind nebulae, the supernova remnant Sgr A east, and stellar wind
shocks (Quataert \& Loeb 2005).  For most of our calculations, we
focus on a source arising near the black hole producing Sgr A*.

Combining the motions induced by magnetic fields with our
proton-proton scattering cross-section, we have developed a
Monte-Carlo technique to determine where protons of a given energy
scatter.  Magnetic fields act as an effective scattering opacity while
proton-proton scattering acts as an absorption term (because we
destroy the protons upon their first scattering).  We first develop a
3-dimensional grid (typically 200 zones on a 6~pc side - 8 million
zones total) and determine our two ``opacities'' at each grid point.
We then launch packets from the center of this region (to mimic a Sgr
A* or stellar wind source) and from a shell at roughly 5~pc (to mimic
a Sgr A East supernova remnant source) and follow their trajectory
until they scatter to produce protons.  The magnetic fields are able
to completely alter the direction of the proton in any zone, and we
determine which adjacent zone the proton travels to by assuming a
diffusion approximation.  The probability the proton moves into a
specific adjacent zone is inversely proportional to the magnetic field
``cross-section'' in that adjacent zone.  In each zone, we calculate
the probability for the proton to scatter and produce pions and, using
a random sampler, determine whether a scattering event has occurred.
We typically run 10 million particles per simulation to obtain enough
statistics to make maps of the projected scattering surface.

The photon emission is proportional to the pion production rate, and
because the pion decay is rapid, the location of the $\gamma$-ray
emission is at the same spot as that of the proton scattering.  Hence
we can use the scattering location to study the distribution of
$\gamma$-rays.  Before we calculate the photon energy distribution, we
focus on the spatial distribution, and hence the scattering location.
Our first step is to study the dependence of this spatial distribution
on the proton energy.  In our first series of plots, we present
contours of the pion production per zone (a 0.06~pc$\times$0.06~pc
square projected area) as a function of the Galactic coordinates.  We
normalize this production rate to the fraction of total protons
launched at that energy.  Figure 6 shows these contours for the wind-blown bubble
density solution for 3 different proton energies: 0.1, 10, and 1000
TeV.  Although the wind-blown bubble region extends beyond 2~pc,
surface area effects produce the largest surface emission to be
centered around the GC.  As the energy of the protons increases, fewer
and fewer pions are produced in the center, due to the decreased
effect of magnetic fields for higher energy photons.

How sensitive is this result to the details of our density profile?
We also studied the effect of a lower density in the surrounding
medium.  Figure 7 shows the results of a simulation where this density
is lowered by an order of magnitude and expanded (as we would expect
from \S\ 4).  The result is fairly insensitive to this change (and
will remain so until we lower the density of the molecular cloud
sufficiently that the particles stream through it (below $\sim 100
{\rm cm^{-3}}$).  In addition, it is possible that the density
profile, like Sgr A East, is offset from the center of Sgr A*, but
since the winds arise from stars orbiting Sgr A*, this offset is
likely to be small.  As an extreme test of this uncertainty, we moved
the center of the density distribution 2~pc away from Sgr A* to study
the effects of such an offset.  The corresponding spatial distribution
of the $\gamma$-ray emission is shown in Figure 8.  Although the shape
is slightly changed, the result is very similar to our simple,
spherically symmetric model centered on Sgr A*.  Even though both of
these alterations show differences from our standard model, it would
be difficult, if not impossible, to observe these differences with
existing telescopes.

If we model the density profile produced by Rockefeller et al. (2004), the
peak densities occur in the torus and right near Sgr A*.  The
resultant contours are far from symmetric (Fig. 9).  The density
structure of the circumnuclear disk is clearly shown in the image.
In this manner, we can use the high energy emission to probe the
density structures of the GC.  This emission provides an
independent means to study these densities perturbations.  The
combined structure, using the Rockefeller et al. (2004) density profile in
the inner 2~pc and our derived wind-blown bubble structure beyond
shows that dense structures such as the circumnuclear disk will still 
produce noticeable variations on the high-energy signature (Fig. 10). 

How sensitive are our results to our choice of opacities?  Fig. 11
shows the pion production distribution assuming that the power index
for the magnetic fields is $q=2$ instead of $q=1.5$.  We show plots
for two scale lengths in the magnetic field: $10^{16}$~cm and
$10^{18}$~cm.  If the magnetic scale length is too short, much of the
enhancement caused by the circumnuclear disk is washed out.  Fig. 12
shows the pion production distribution with the full equation for the
proton-proton scattering for 3 energies: 0.1~TeV, 10~TeV, and 1000~TeV.
Over this range, the proton-proton scattering cross-section varies by
less than a factor of 2 so it is not surprising that it does not change
the distribution results considerably.  These modifications will,
however, have a larger effect on our gamma-ray spectra.

Finally, can we distinguish a Sgr A East source from a more centrally
located source.  In all of our central source calculations, the
dominant signal occurs in a projected 2~pc circle around Sgr A*.  But
if the source is the supernova remnant, the source will be extended
(Fig. 13).  Unfortunately, the distribution is still limited to a
$\sim 5$~pc radial emission source in the GC, currently
below the distribution resolution from HESS (Aharonian et al. 2004).
If the density enhancement of Sgr A East truly is offset from Sgr A*,
there is a possibility that we can distinguish this source from that
of a Sgr A* source with the current data (we discuss this further
below).  Clearly, improved data will be able to distinguish these two
sources.

A summary of our simulations is given in Table 1.

\section{The High Energy Photon Emission}
\label{sec:spect}

The plots from section~\ref{sec:protons} show the spatial distribution 
of the position where our protons scatter to produce pions for a 
set of different density profiles.  Since the pions decay in less 
than $10^{-10}~$s, this is also the spatial distribution of the 
high-energy photons.  Another way to understand the physics and the 
physical sources of this emission is to study the spectra produced by 
the different simulations.  If we want to calculate the spectrum of 
this emission, we must estimate the energy distribution of pions 
and the resulting energy distribution of high energy photons.  

For the energy distribution of pions, we assume a differential
cross-section [$d\sigma(E_p,E_{\pi^0})/dE_{\pi^0}$] for the proton
scattering (Blasi \& Colafrancesco 1999, Fatuzzo \& Melia 2004):
\begin{equation}
d\sigma(E_p,E_{\pi^0})/dE_{\pi^0} = \sigma_0
f(E_{\pi^0}/E_p)/E_{\pi^0}, 
\label{eq:pidist}
\end{equation}
where $E_p$ and $E_{\pi^0}$ are the proton and pion energies,
respectively, $\sigma_0$ is the cross-section for proton scattering we
used in in section \ref{sec:protons}, and $f_{\pi^0}(E_{\pi^0}/E_p)$
is
\begin{equation}
f(E_{\pi^0}/E_p)=0.67(1-E_{\pi^0}/E_p)^{3.5} + 0.5 e^{-18E_{\pi^0}/E_p}.
\end{equation}
If we assume momentum and energy conservation and that two photons
dominate the decay, the energy distribution of the $\gamma$-ray
photons we observe is flat between the limits of
$(E_{\pi^0}-P_{\pi^0})/2$ and $(E_{\pi^0}+P_{\pi^0})/2$ where
$P_{\pi^0}$ is the pion momentum.  Combining these two distributions,
we derive a spectrum for the $\gamma$-rays emerging from proton
scattering.  Note that the nature of equation~\ref{eq:dist} means 
that the gamma-ray spectrum need not be identical to the spectrum 
of emitted protons, but, as we shall see, these two spectra are 
fairly similar.

To obtain an observed spectrum of $\gamma$-rays, we must first assume
an injected spectrum of protons.  For our standard model, we assume
that the proton energy distributions falls off as $E_p^{-2.3}$.  We
include the emission protons from 10~GeV up to 1000~TeV.  The exact
value for this power index is fairly uncertain (we discuss the
dependence on this index below).  Our standard model also consists of
our wind-blown bubble condition (assuming a $10^4 {\rm cm^{-3}}$
surrounding density) with the circumnuclear disk added in the central
2~pc.  The total spectrum within 6\,pc is shown in
Fig.\ref{fig:specsuite}.  To explain the flux of $\gamma$-rays
observed by HESS from 165~GeV to 10~TeV, we need a total energy in
high-energy $\gamma$-rays of $\sim 1.4\times 10^{35} ~ {\rm erg
s^{-1}}$.  For our simulation, such a flux would require a proton flux
above 200~GeV of nearly $5\times10^{37}~{\rm erg s^{-1}}$.  Such
values are easily within the possible proton fluxes predicted by
models accelerating protons near the Sgr A* black hole (Liu et
al. 2006).

Although between 10 and 100 GeV, the spectral index is close to that
observed by HESS ($\Gamma \sim -2.2$), in the actual HESS regime
(above 0.1~TeV), the spectral index is generally a bit steeper (below
-2.2).  This index depends sensitively on our assumptions of the
proton energy distribution, the density profile, and the opacities.
In Fermi shock acceleration models, we are restricted to source energy
distributions of the protons with an index near 2.  Not so with the
stochastic model of Liu et al. (2006).  If we assume the proton energy
distribution falls off as $E_p^{-3.3}$, we obtain the dashed curve in
Fig.~\ref{fig:specsuite}.  Here the spectral index of the
$\gamma$-rays is much steeper than the observed rate.  Whereas 
shock acceleration models predict roughly the correct energy distribution,
the stochastic acceleration model must come up with an argument for
getting an answer similar to the shock accleration models.  For this
simulation, we require a proton emission energy above 0.2~TeV of
$10^{37}~{\rm erg s^{-1}}$ to match the HESS flux.  The spectral index
is also sensitive to the density distribution.  Using the wind-blown
bubble profile alone (with the proton distribution of $E_p^{-2.3}$),
we obtain the dotted curve in Fig.~\ref{fig:specsuite}.  The spectral
index for this simulation (using a proton spectral index of -2.3) is
steeper than our corresponding combined simulation (solid curve,
Fig.~\ref{fig:specsuite}), but still within an acceptable range to fit
the data.  For this model, the proton flux required above 0.2~TeV
would be below $10^{37}~{\rm erg s^{-1}}$.  Similarly, we have modeled
both a supernova remnant spectrum (dot-dashed) and altered opacity
calculation (long-dashed).  All fit the observations comparably well.

If we had spatial information of the spectrum, we would have another
set of constraints on our model.  Figure~\ref{fig:protspat} shows the
energy distribution of our protons as a function of the position at
which they scatter from 0.24~pc to 2.4~pc for our standard model
(bottom) and our $q=2$ magnetic field distribution model (top).  The
slope does not change much, but the data gets much noiser as we move
out from the center and get fewer scattering events (especially at the
high energy range).  The decrease in high-energy scatters will alter
the slope of the gamma-ray emission.  Figure~\ref{fig:specspat} shows
the spectrum of the $\gamma$-rays as we move out from 0.24~pc to
2.4~pc for these same two simulations.  The spectral index in the
0.1-10~TeV range is approximately -1.8 close to the black hole, but
falls to -3.8 if we look at material beyond 2.4~pc for our standard
model.  This is because so few high-energy protons scatter beyond the
inner 0.5~pc of Sgr A*.  Of course, this result depends sensitively on
both our proton spectral index, our density distribution, and even on
our choice for the power law index for the scale of our magnetic
fields.

What can we take away from these calculations?  Within the
uncertainties in our proton emission, proton scattering, density
profile, etc., we can easily match the observations of both
$\gamma$-ray spectra and flux.  This is simply because we have several
large uncertainties that can be tuned to fit the results, and only 2
pieces of data.  But, if we had more detailed information, such as the
evolution of the spectral index with time, we could place stronger
constraints on our models.  If we had, in addition, a detailed spatial
map of the emission, we could further constrain our models and
ultimately get a better understanding of the magnetic field
distribution and the energy distribution of our cosmic ray protons
produced by our acceleration mechanism.

\section{Conclusions and Future Prospects}

A wind-blown bubble density structure provides an ideal first-order
approximation of the $\sim 5$~pc region surrounding Sgr A*.  This
structure is not only a prediction of theory, but it also is supported
by several observations.  It allows a more reasonable estimate of the
supernova explosion energy and explains the X-ray ridge observed near
Sgr A*.  With a single parameter (the density of the medium beyond the
bubble), we can predict the size of the bubble.  Bear in mind,
however, that this model is still only a first-order approximation;
many additional molecular structures complicate this region.

We apply this simple density structure to a specific model for the
source of high-energy $\gamma$-rays near the GC: scattering of
high-energy protons.  Given the uncertainties in the source (both its
energy and its spatial distribution), the opacities (especially the
effects of magnetic fields), and the densities in the GC
region, it is not surprising that we can construct multiple solutions
that match the current observations.  But our calculation does allow
us to determine which uncertainties must be better understood to
constrain the models further with the existing data.  In addition, we
can determine which future observations will most help us constrain
the models.

We have shown that detailed spatial maps can help us study the density
structure in the GC region, but if the source of protons
is indeed Sgr A*, the resolution required is far beyond what we can do
with current telescopes.  What the observations can constrain is the
source itself.  If Sgr A East were the source, the emission would be
quite extended.  Figure 17 shows the normalized flux as a function of
radius.  For a source near Sgr A*, the gamma-ray emission really is
point-like, within the inner 2~pc.  But a Sgr A East supernova remnant
source argues for an extended source of gamma-ray emission (out to
5~pc) that increases out to 5~pc and then drops precipitously.  The
innermost two pixels in the current data correspond roughly to 7\,pc
and our emission profile from Sgr A East is just barely consistent
with these two pixels.  The strong drop in emission at 5\,pc would
mean that a pixel centered at 7\,pc should be a factor of 2 lower than
the peak.  As data improves, this emission source might be easily
ruled out.  But remember that Sgr A East is not a 5\,pc spherical shell.  
If the emission is dominated by the shorter axis of Sgr A East, this
emission would center around 3~pc instead of 5~pc.

The centroid of the emission also places constraints on a Sgr A East
source.  Sgr A East is offset by 2~pc from Sgr A*.  If the centroid
truly is exactly on Sgr A *, and the density profile of Sgr A East
truly is offset (not just the emission), we can also rule out Sgr A
East.  Note that for a Sgr A* source, even if the termination shock 
of the wind-blown bubble is offset by 2\,pc, we found that the centroid 
of the high energy emission does not move by more than 0.1\,pc.

Finally the question arises whether the extended gamma-ray emission
can be used as a constraint on the source.  If we want to claim that
the extended emission is produced by the same proton source as the
point-like source, we can use the observations to constrain the proton
source.\footnote{Note, however, that the only reason to assume this is
because the spectra are similar---but if we use shock acceleration to
produce our high-energy protons, we would expect any source to produce
roughly the same spectrum.}  For the Sgr A East source, the primary
constraint arises because, given the new estimate for the age of the
remnant by Rockefeller et al. (2006), Sgr A East has been producing
high-energy protons for at most 1000~y.  For those protons to emit
gamma-rays 100~pc away, there must be a ``hole'' where the density is
low and the scale-length of the magnetic field is high.  The proton
travel time is this distance (100\,pc) times the effective optical
depth of the protons divided by the speed of light.  To make this
timescale fall below 1000\,y, this roughly requires an optical depth
of less than 2-3, implying that the scale-lenght of the magnetic field
is greater $\sim$ 50~pc.  This timescale constraint is not important
for a continuous source like Sgr A*.  However, the fact that the
spectral index is nearly identical suggests that either the
opacities---in particular, the magnetic field effective opacity---must
be energy independent or that a similar hole exists.

We can not absolutely rule out a Sgr A East source for the high energy
emission, but taking all of the above constraints together, we have to
introduce a number of caveats to make a Sgr A East source match the
existing data.  With accurate density maps and a precise localization
of the high-energy source, it is likely that this source can be ruled 
out in a more absolute sense.

Finally, the spectral distribution of the gamma-ray emission depends on all of
our uncertainties: source, opacities, and density distribution in the
Galactic center.  However, the spectral distribution as a function of
space is an ideal probe of our opacities, in particular the
distribution of the magnetic fields (compare the plots in Figure 16).

{\bf Acknowledgments} This work was carried out under the auspices of
the National Nuclear Security Administration of the U.S. Department of
Energy at Los Alamos National Laboratory under Contract
No. DE-AC52-06NA25396.

{}

\newpage

\begin{deluxetable}{cccc}
\tablewidth{0pt}
\tablecaption{Distribution Calculations\label{tab1}}
\tablehead{
  \colhead{Source}
& \colhead{Density}
& \colhead{Opacity}
& \colhead{Figure}  
}
\startdata

Sgr A* & Wind & Eq.~\ref{eq:tausc-q1.5},\ref{eq:ppflat} & \ref{fig:plot}\\ 
Sgr A* & Wind, Low Dens. & Eq.~\ref{eq:tausc-q1.5},\ref{eq:ppflat} & \ref{fig:plot2} \\
Sgr A* & Wind, Offset & Eq.~\ref{eq:tausc-q1.5},\ref{eq:ppflat} & \ref{fig:plot3}\\ 
Sgr A* & Torus & Eq.~\ref{eq:tausc-q1.5},\ref{eq:ppflat} & \ref{fig:plot4}\\ 
Sgr A* & Torus+Wind & Eq.~\ref{eq:tausc-q1.5},\ref{eq:ppflat} & \ref{fig:plot5}\\ 
Sgr A* & Torus+Wind & Eq.~\ref{eq:tausc-q2},\ref{eq:ppflat} & \ref{fig:plot6}\\ 
Sgr A* & Torus+Wind & Eq.~\ref{eq:tausc-q1.5},\ref{eq:pp} & \ref{fig:plot7}\\ 
Sgr A East & Torus+Wind & Eq.~\ref{eq:tausc-q1.5},\ref{eq:pp} & \ref{fig:plot8}\\ 

\enddata
\end{deluxetable}

\newpage

\clearpage

\clearpage

\clearpage

\clearpage

\clearpage

\begin{figure}
\plottwo{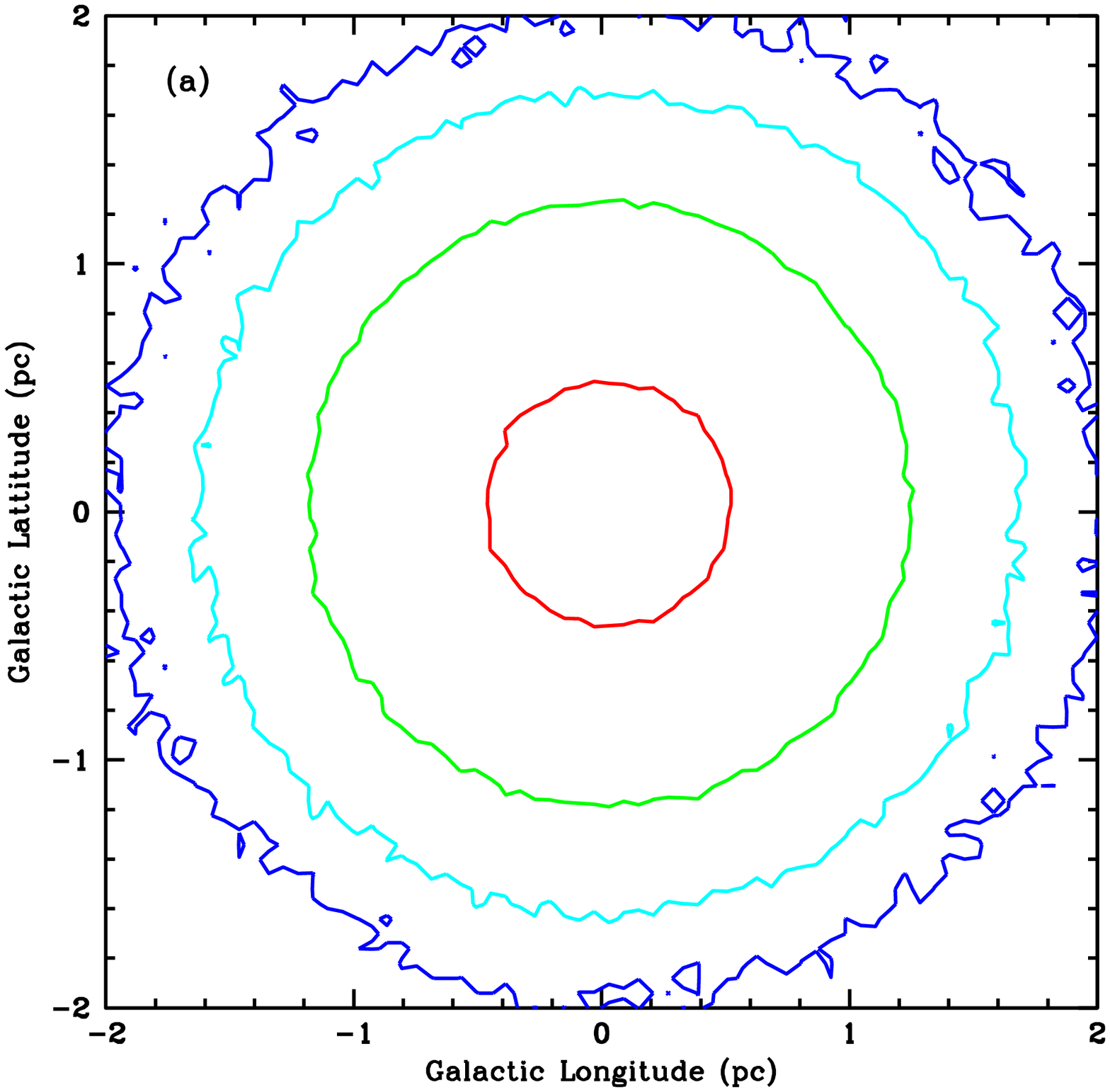}{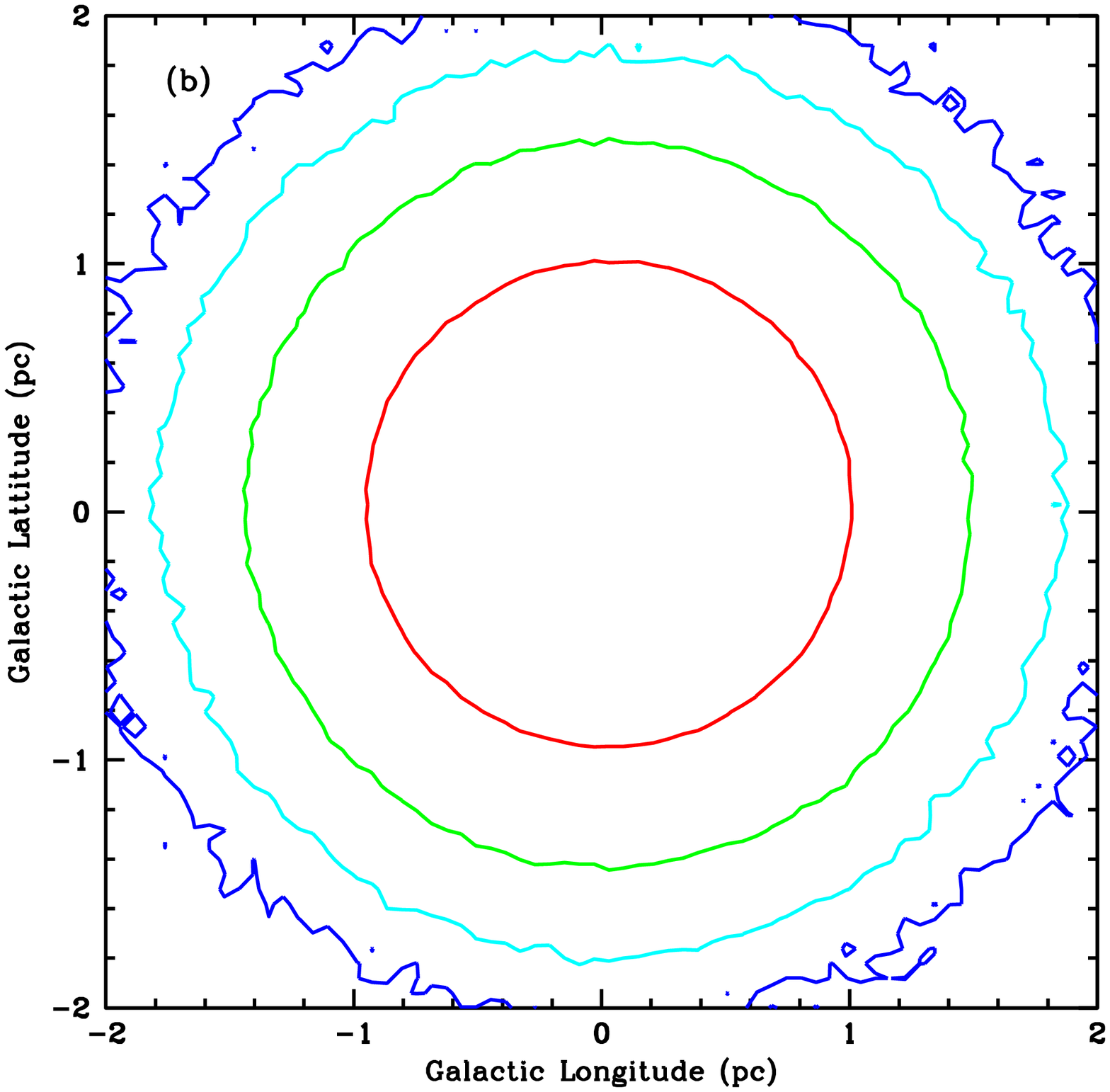}
\epsscale{0.5}\plotone{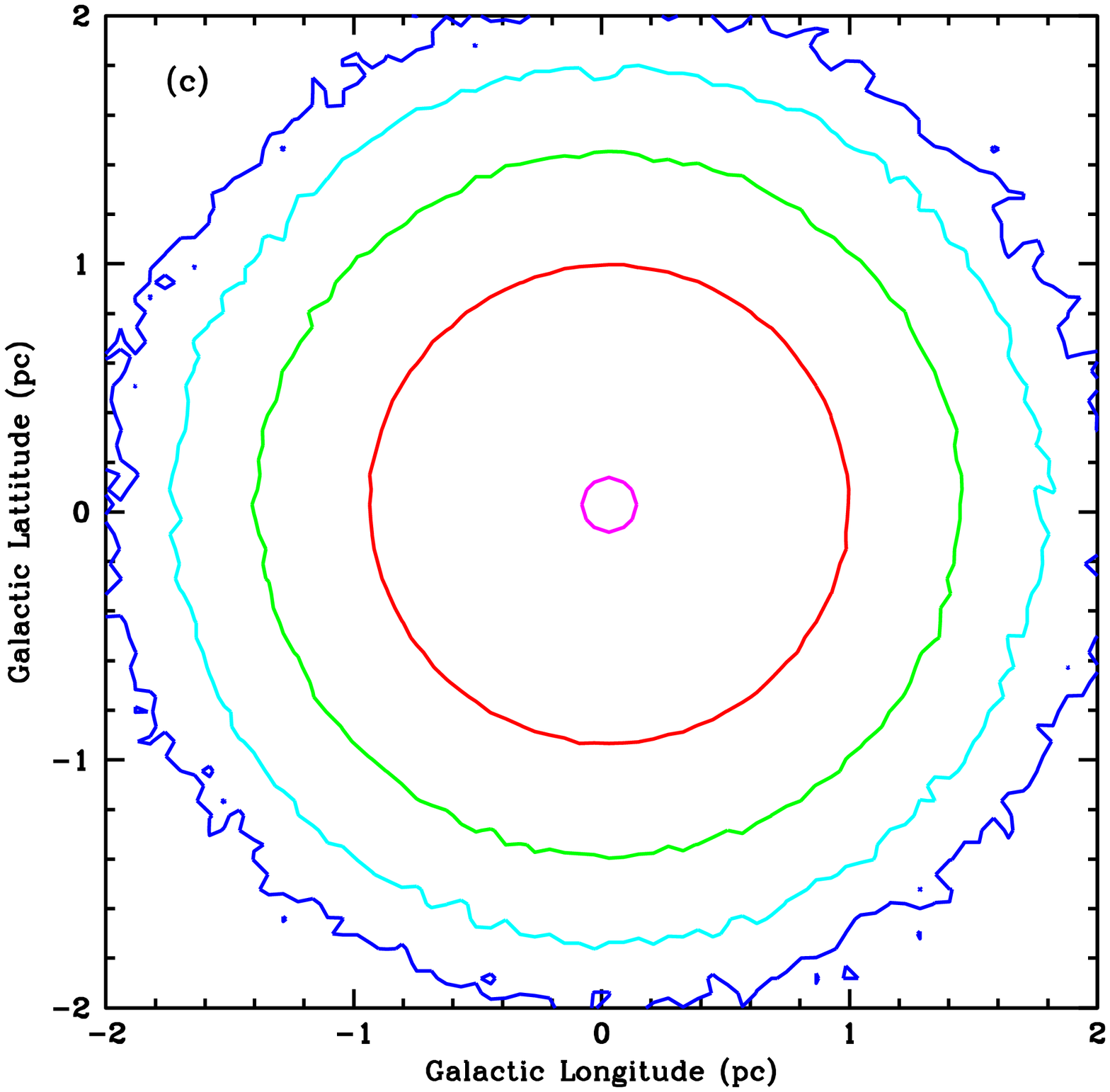}
\epsscale{1.0}
\caption{Contours of pion production as a function of projected
surface area using our standard wind initial conditions: $\dot{M}_{\rm
wind} = 3\times10^{-3}~{\rm M_\odot ~ y^{-1}}$, $v_{\rm wind} = 1000
{\rm km s^{-1}}$ and $\rho_{\rm MC} = 10^4 {\rm cm^{-3}}$.  Each
contour corresponds to the number of total pions produced in each
column (we sum along the projected area to get the total ``observed''
production rate in each 0.06$\times$0.06~{\rm pc}$^2$ zone.  The contour
levels correspond to fractions of: $3\times10^{-3}$ (magneta),
$3\times10^{-4}$ (red), $3\times10^{-5}$ (green), $3\times10^{-6}$
(cyan), $3\times10^{-7}$ (blue).  The 3 plots correspond to (a)
1000~TeV, (b) 10~TeV, and (c) 0.1~TeV protons.  Note that although
the wind density is very extended, the source of pions is actually
quite compact.}
\label{fig:plot}
\end{figure}
\clearpage

\begin{figure}
\plottwo{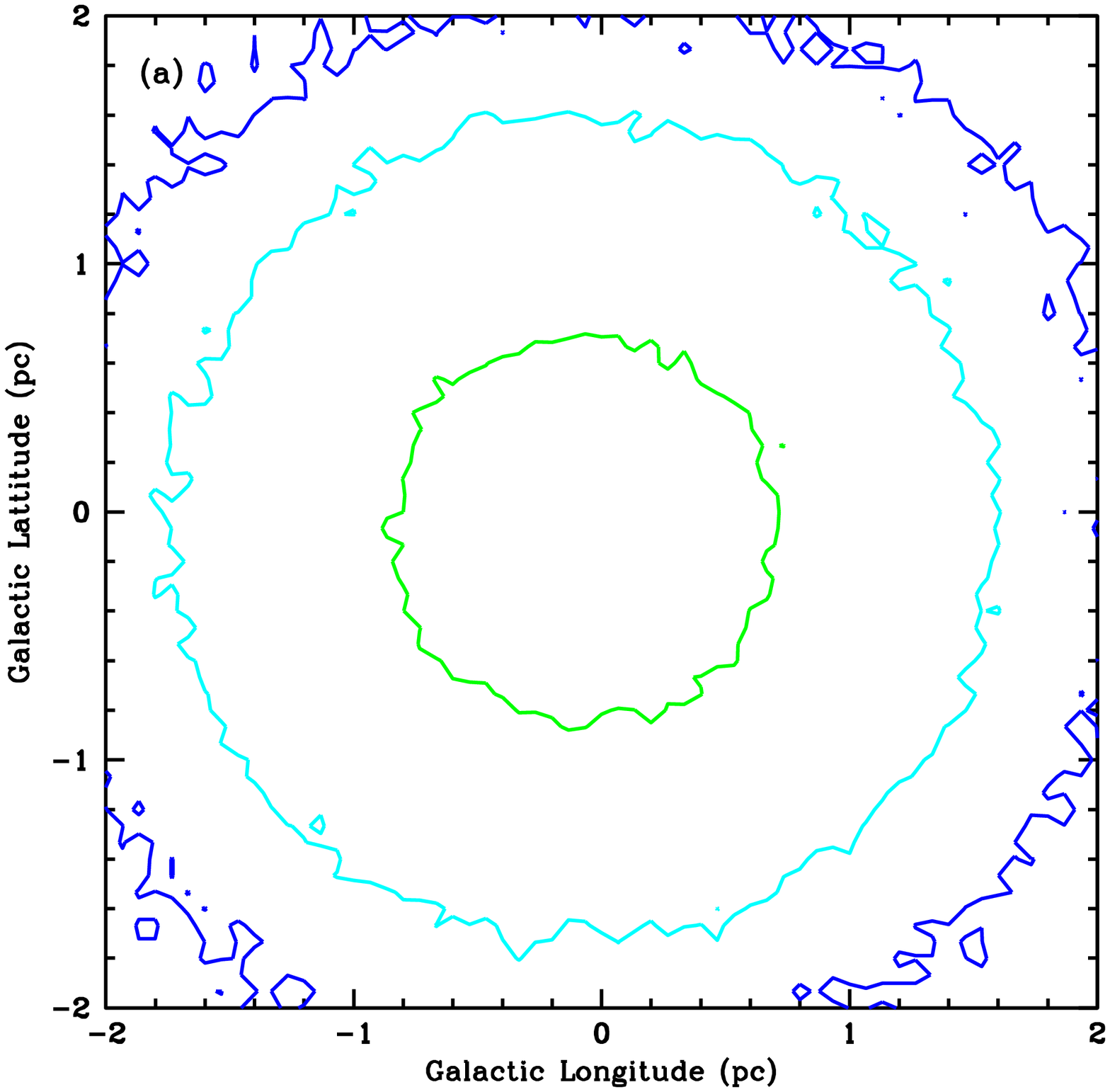}{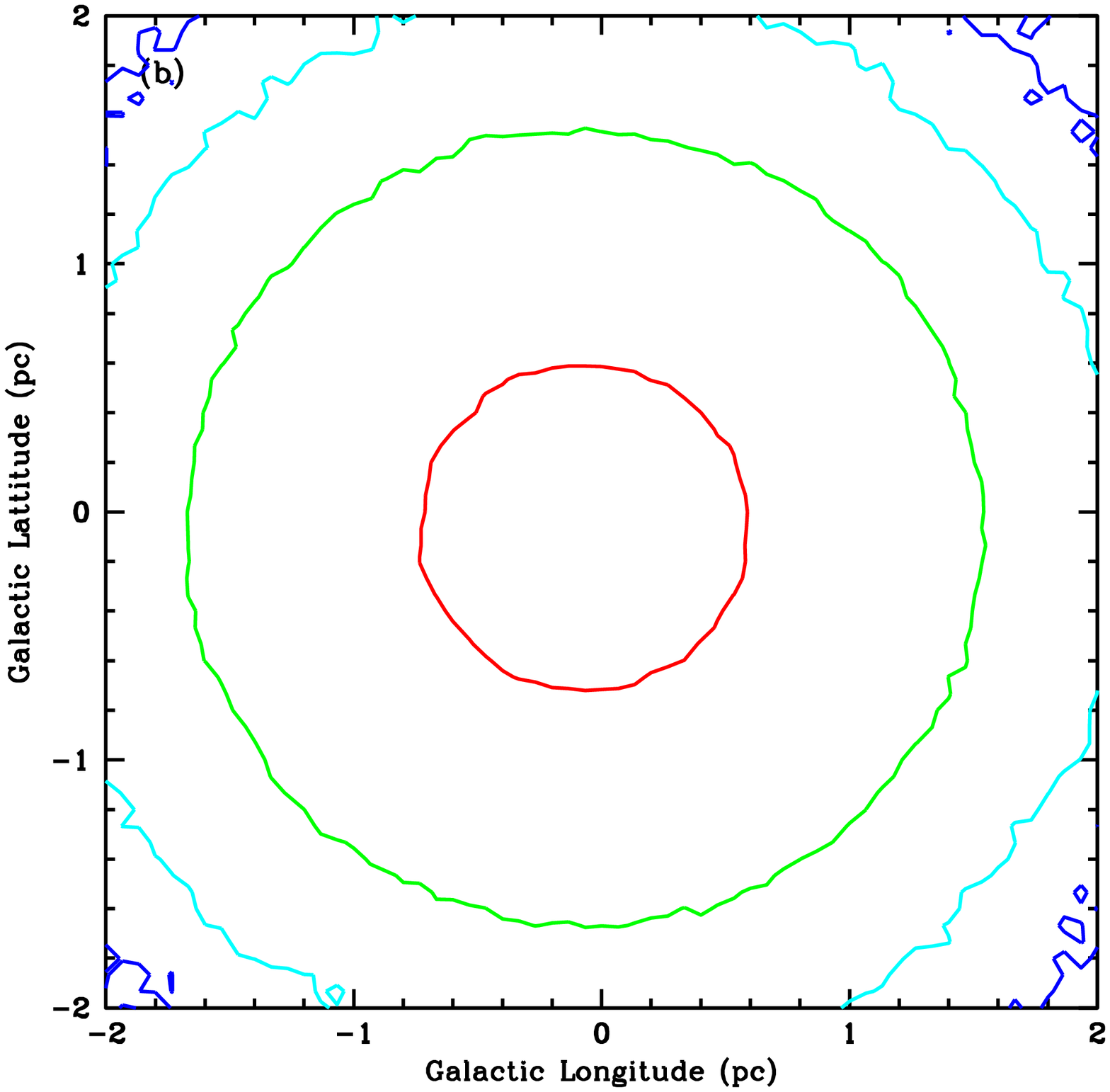}
\epsscale{0.5}\plotone{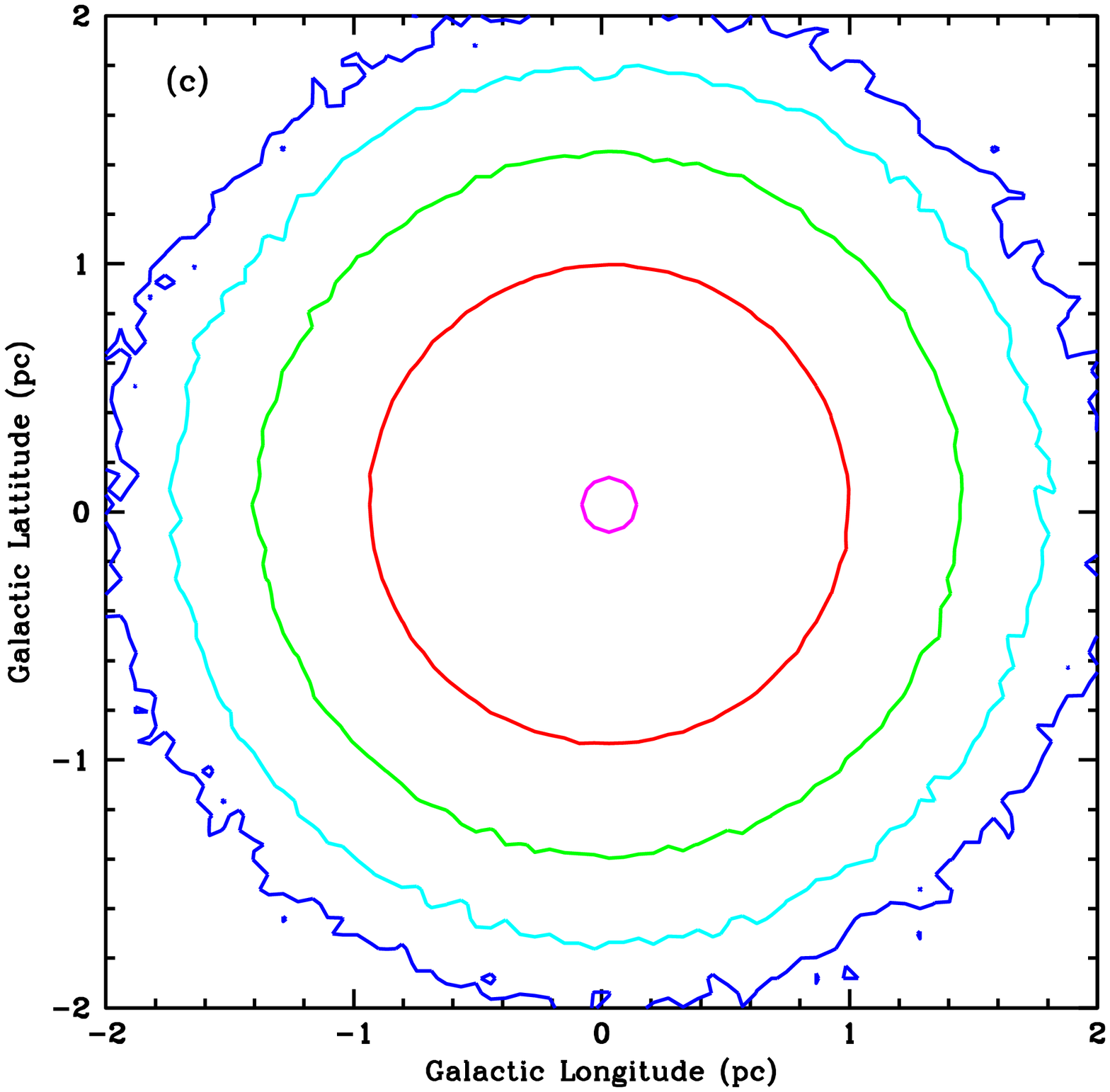}
\epsscale{1.0}
\caption{Same as Fig. 6 but with the density lowered by an order 
of magnitude and the extant of the bubble increased by $10^{0.2}$ 
following the calculation in equation~\ref{eq:rext}.}
\label{fig:plot2}
\end{figure}
\clearpage

\begin{figure}
\plottwo{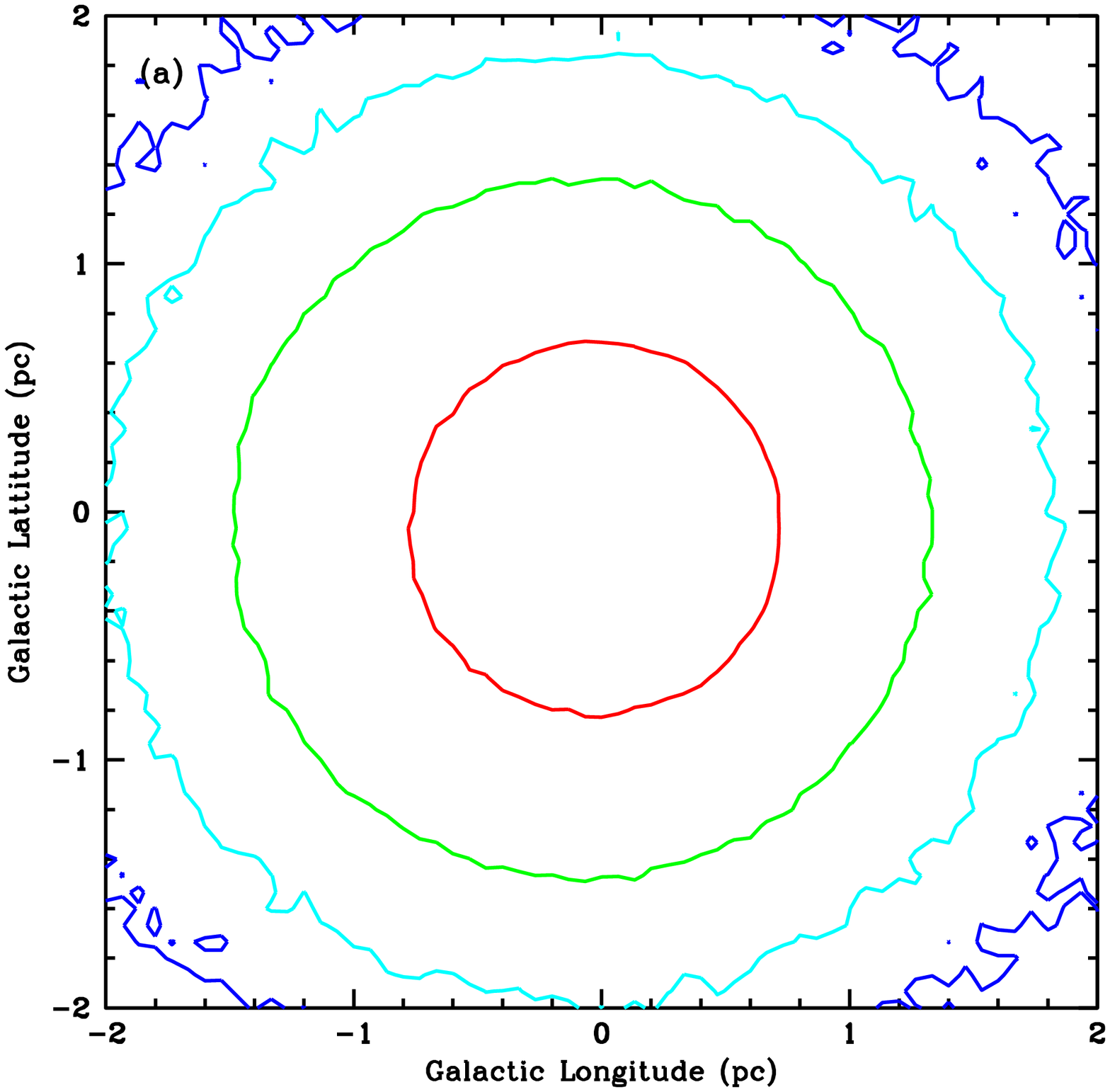}{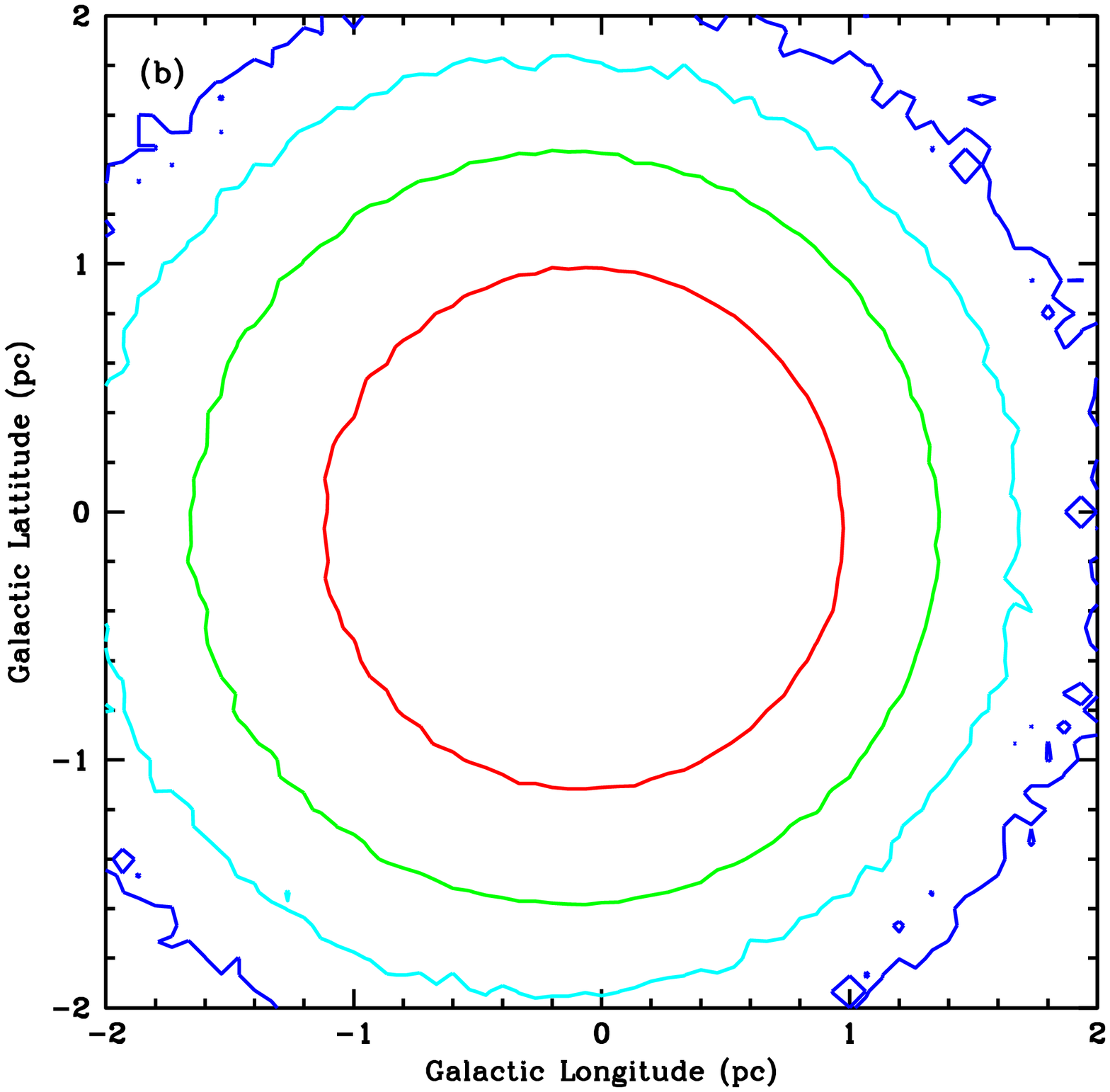}
\epsscale{0.5}\plotone{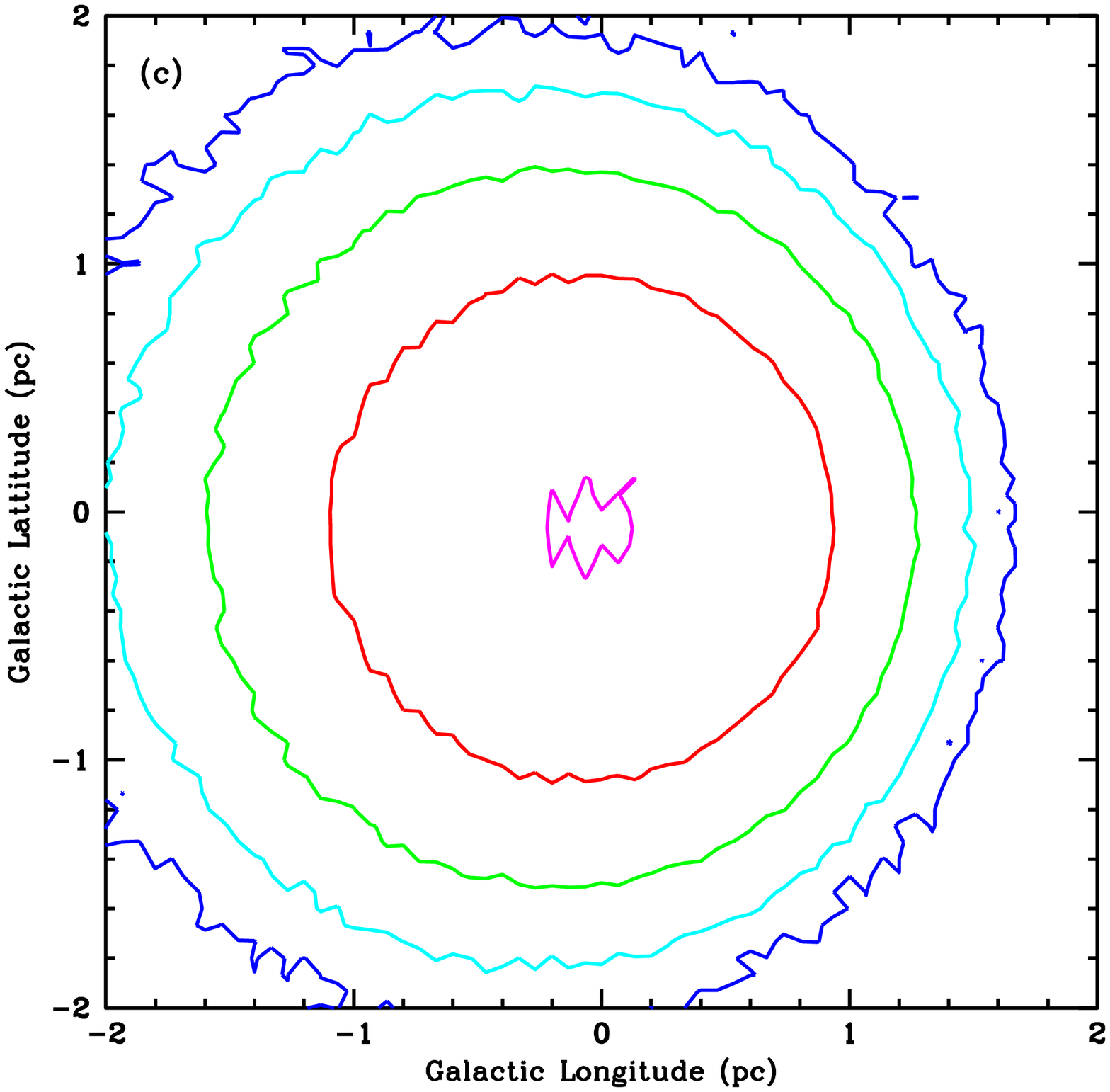}
\epsscale{1.0}
\caption{Same as Fig. 6 but with the density profile moved 2~pc 
off the center of the Galaxy.}
\label{fig:plot3}
\end{figure}
\clearpage

\begin{figure}
\plottwo{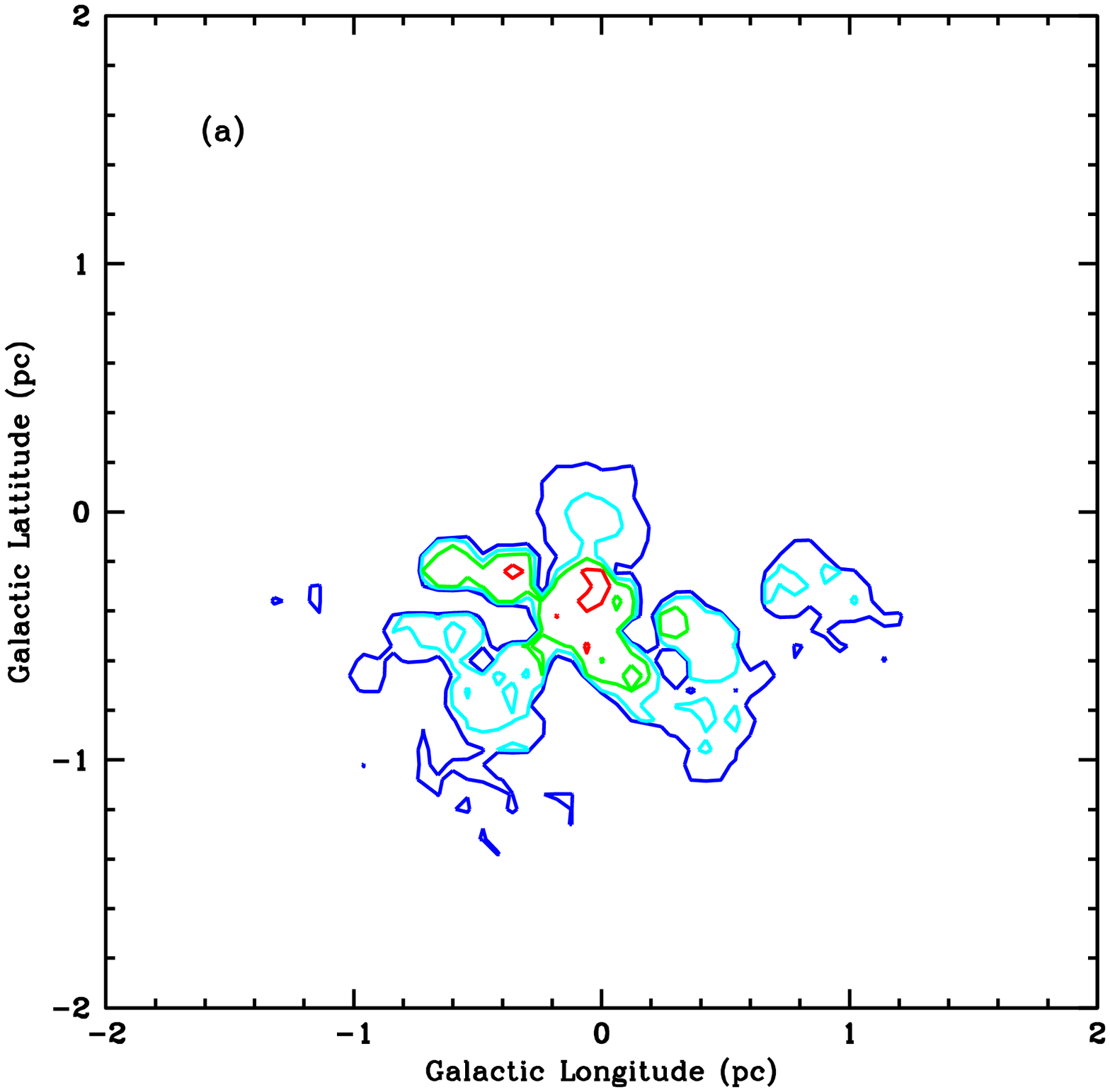}{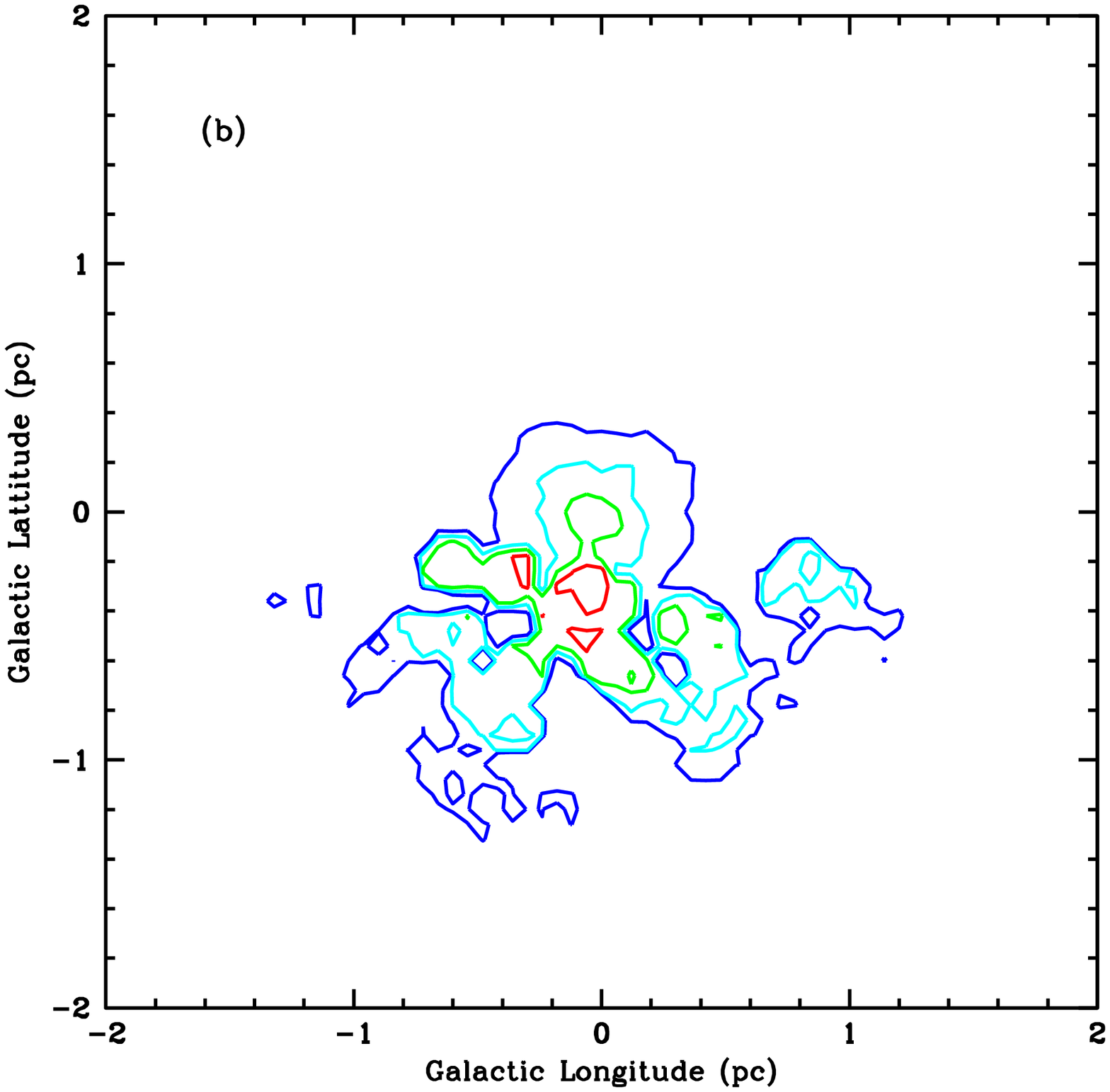}
\epsscale{0.5}\plotone{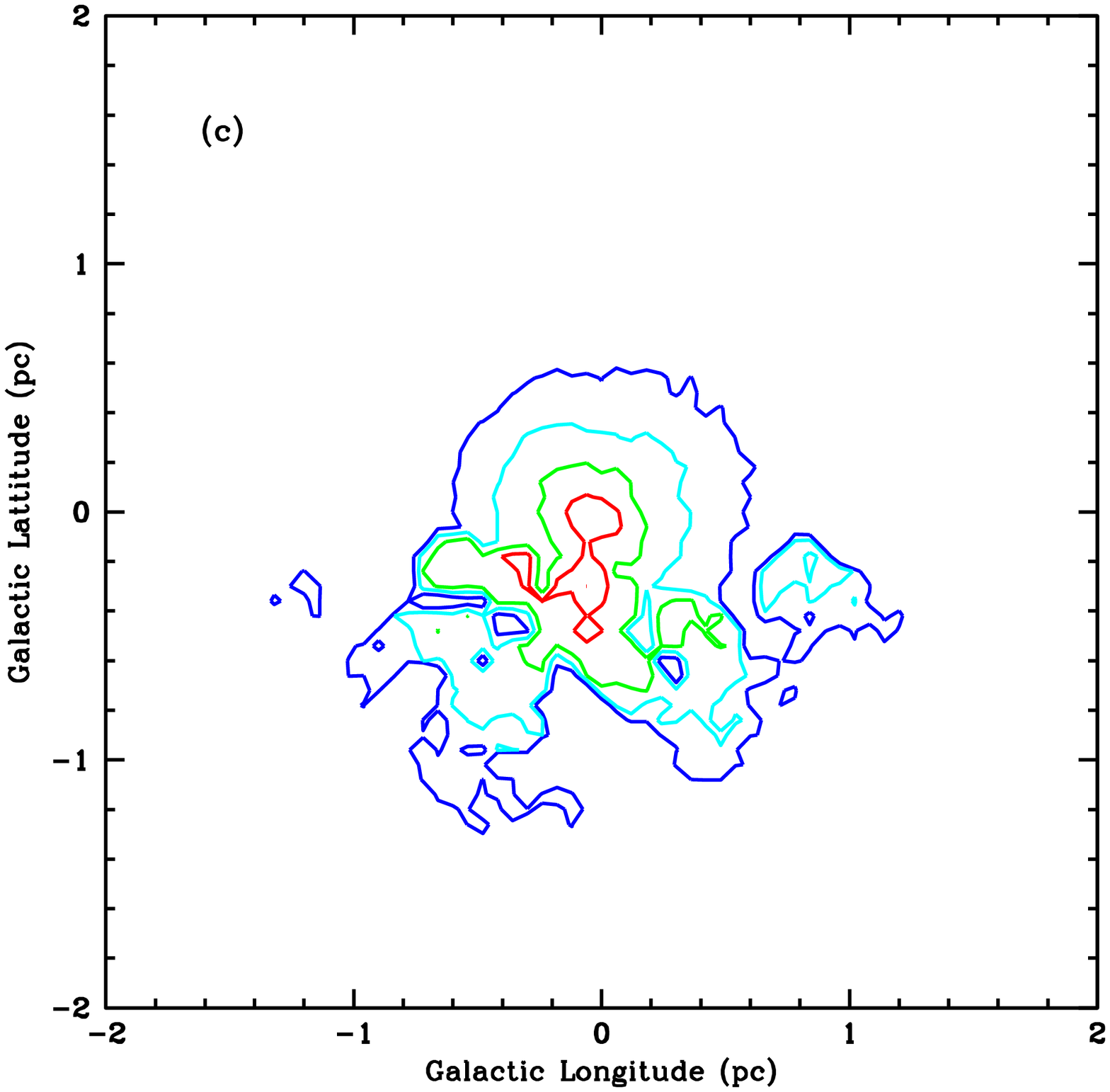}
\epsscale{1.0}
\caption{Same as Fig. 6 but with the density profile set by 
Rockefeller et al. (2004) that is dominated by the circumnuclear 
disk.}
\label{fig:plot4}
\end{figure}
\clearpage

\begin{figure}
\plottwo{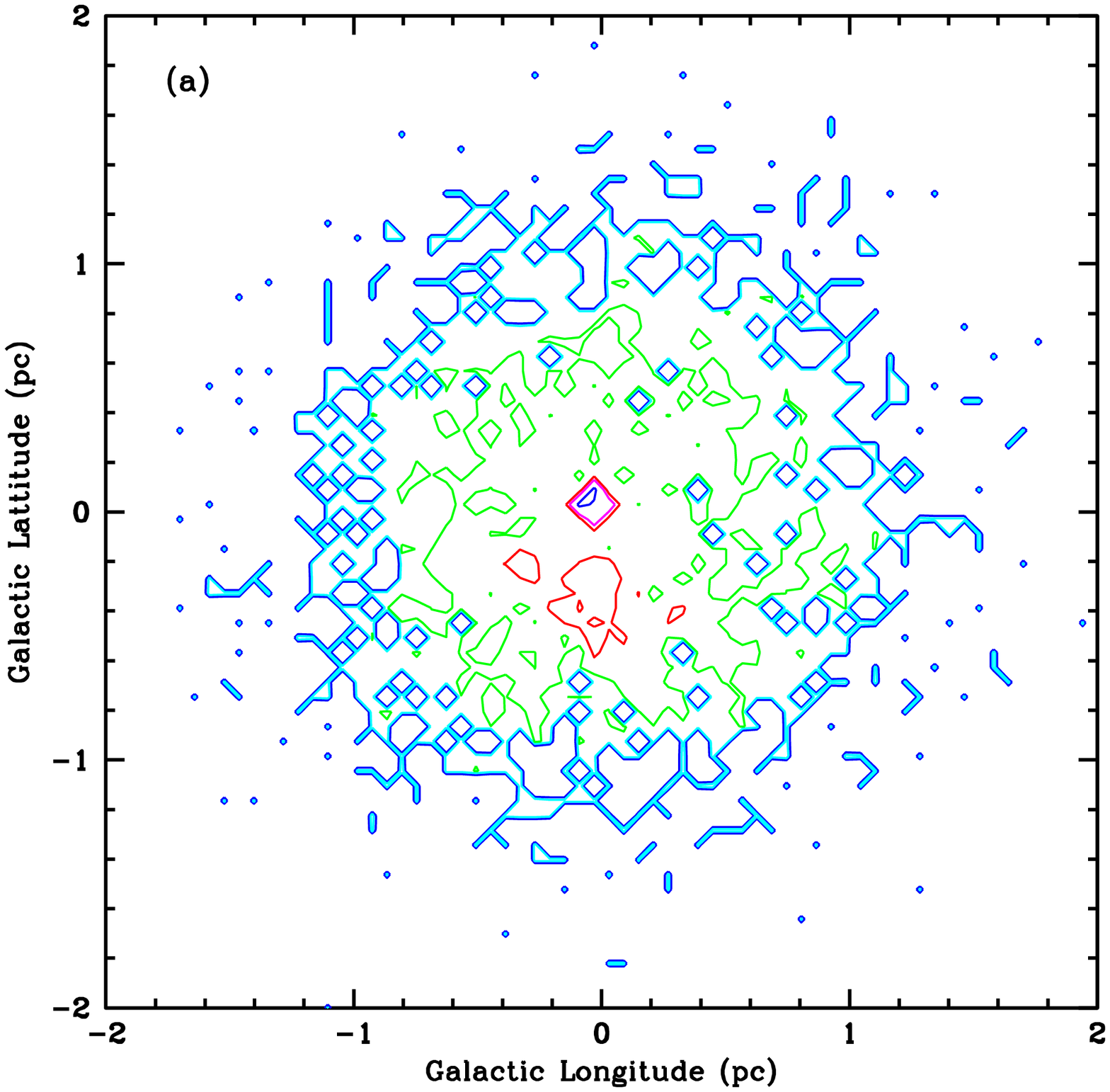}{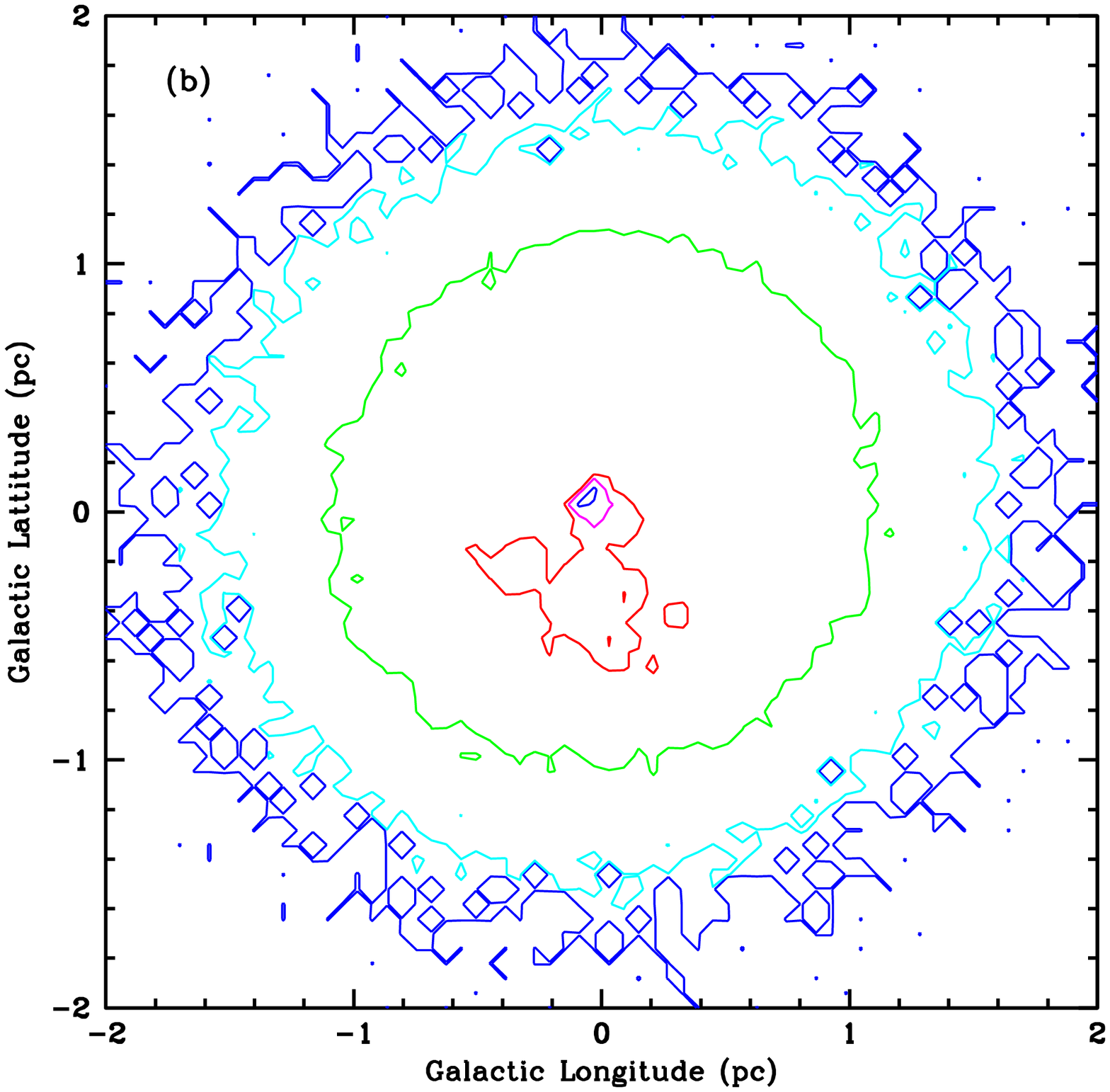}
\epsscale{0.5}\plotone{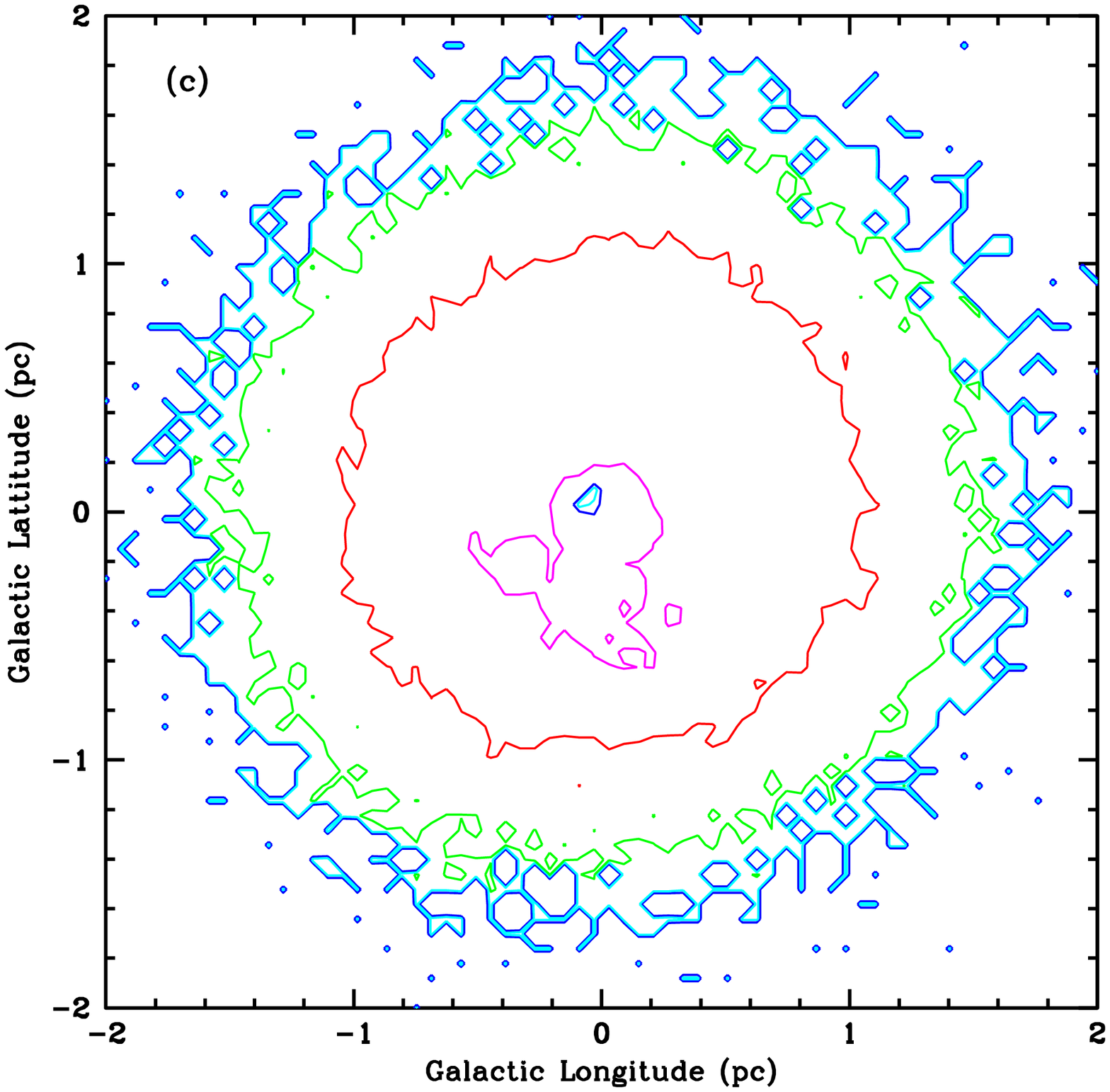}
\epsscale{1.0}
\caption{Same as Fig. 6, but a composite of both our wind profile 
and the inner density set by Rockefeller et al. (2004).  This is 
our standard density profile for studying spectral features in 
\S~\ref{sec:spect}.}
\label{fig:plot5}
\end{figure}
\clearpage

\begin{figure}
\plottwo{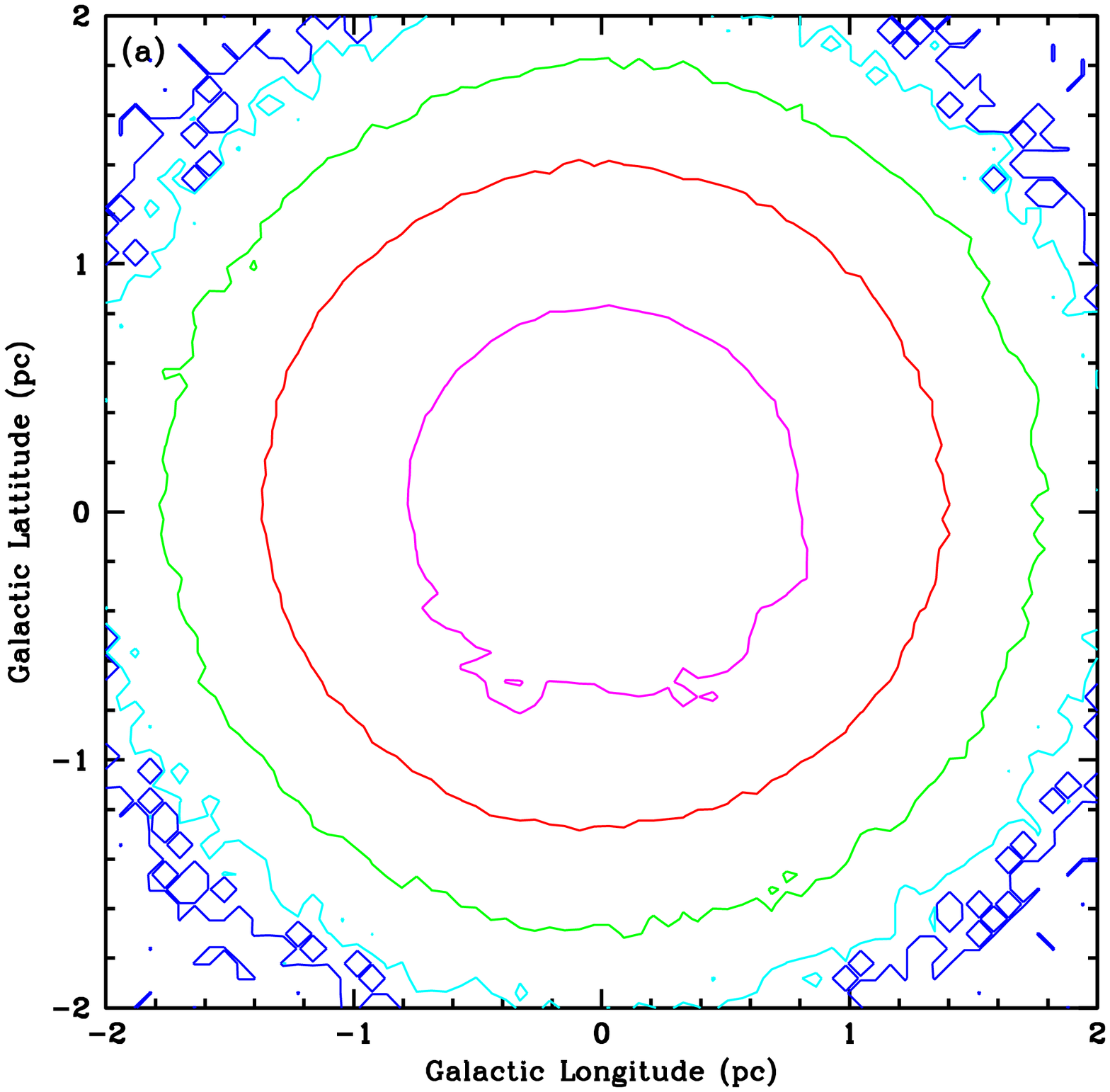}{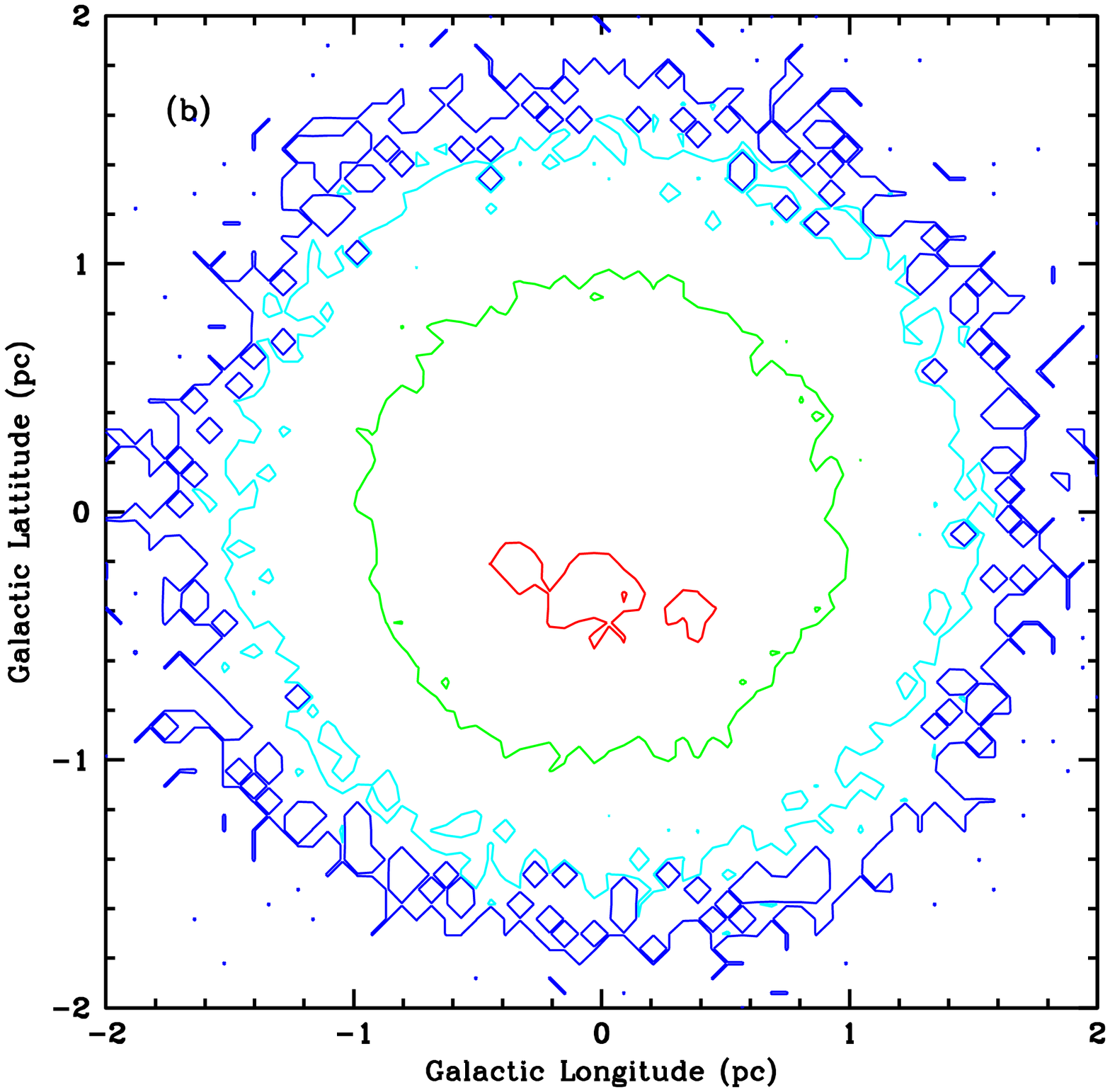}
\caption{Same as Fig. 6, but a composite of both our wind profile 
and the inner density set by Rockefeller et al. (2004) using a $q$=2 
value for the magnetic field distribution.}
\label{fig:plot6}
\end{figure}
\clearpage

\begin{figure}
\plottwo{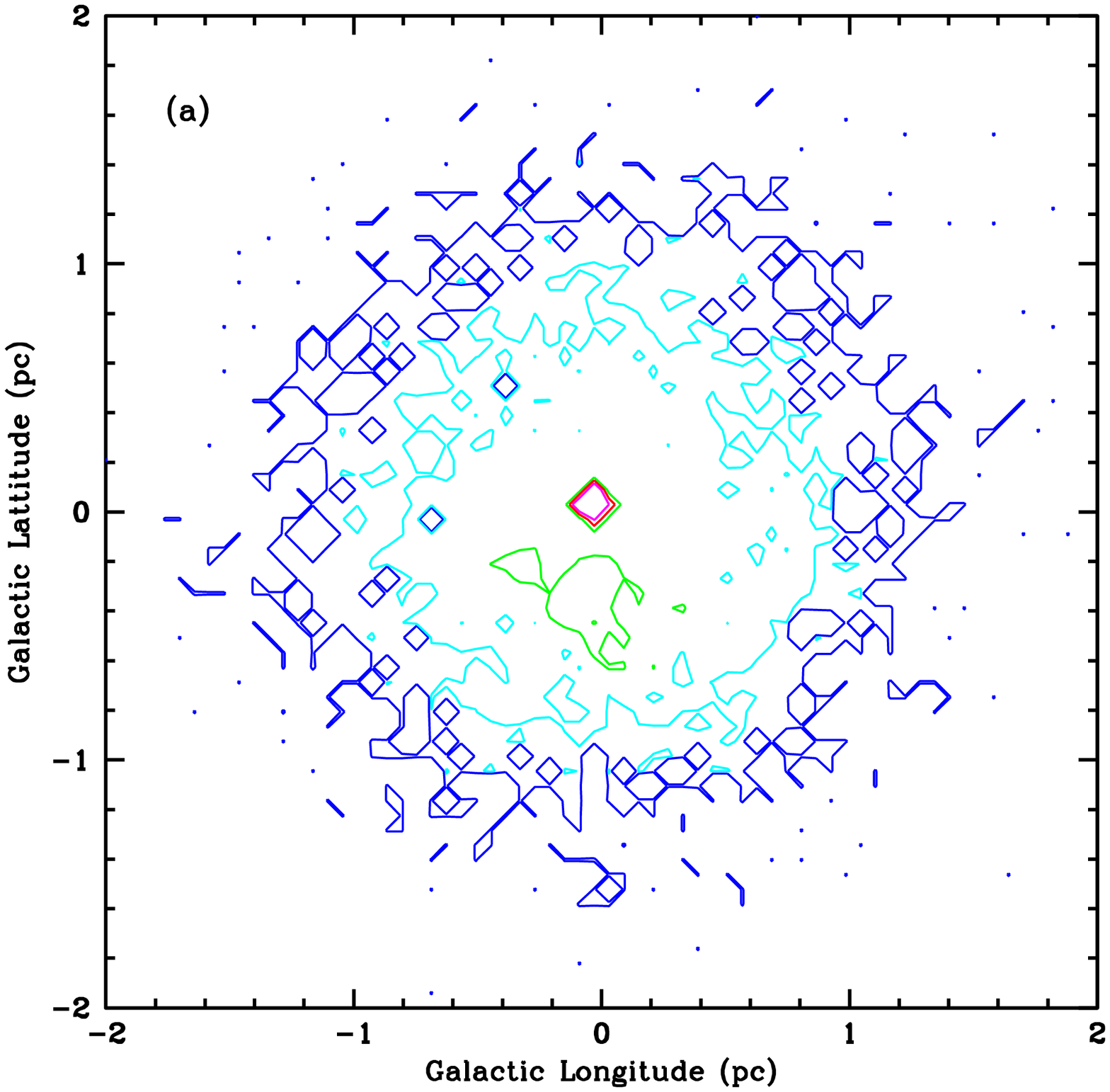}{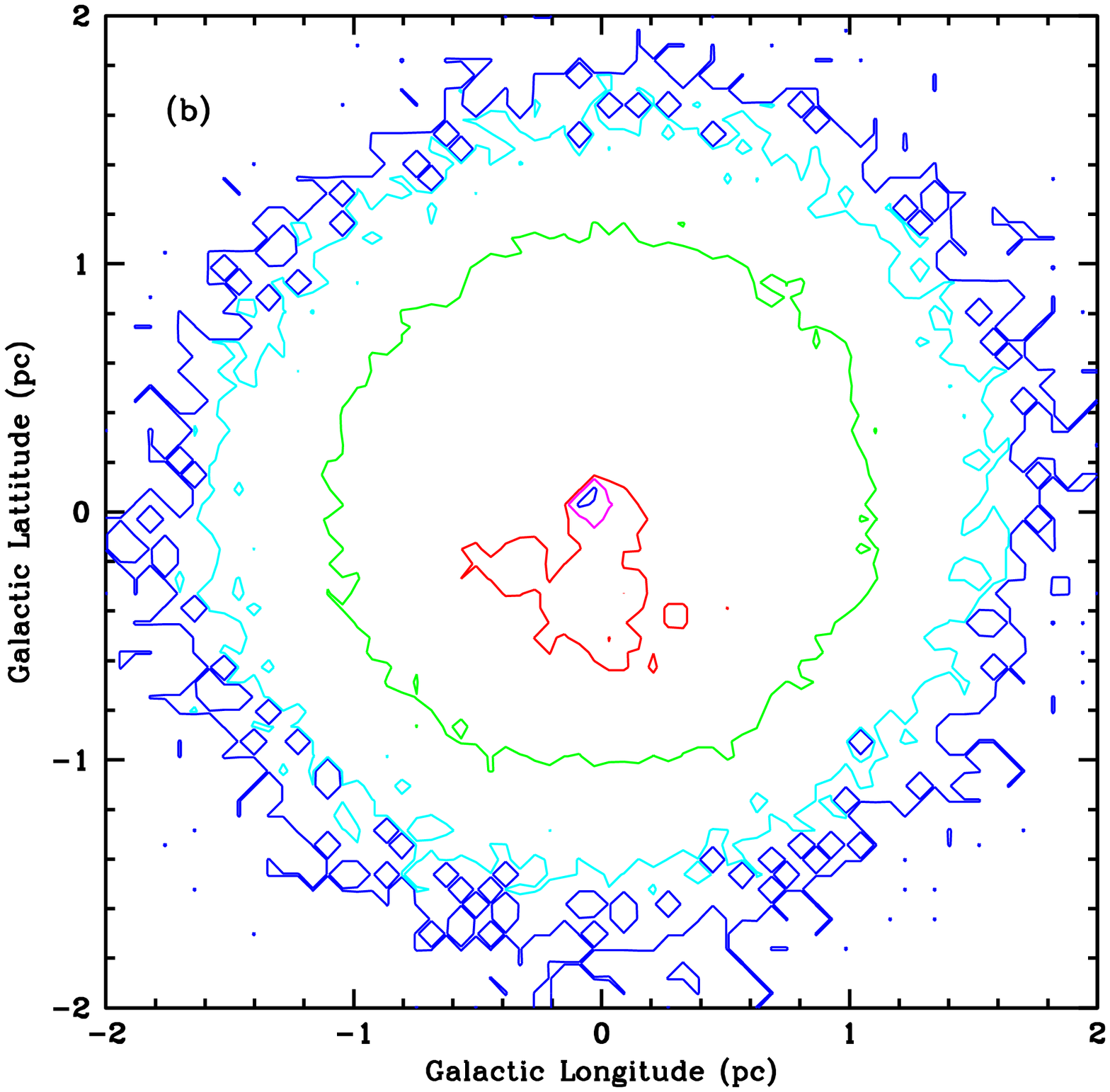}
\epsscale{0.5}\plotone{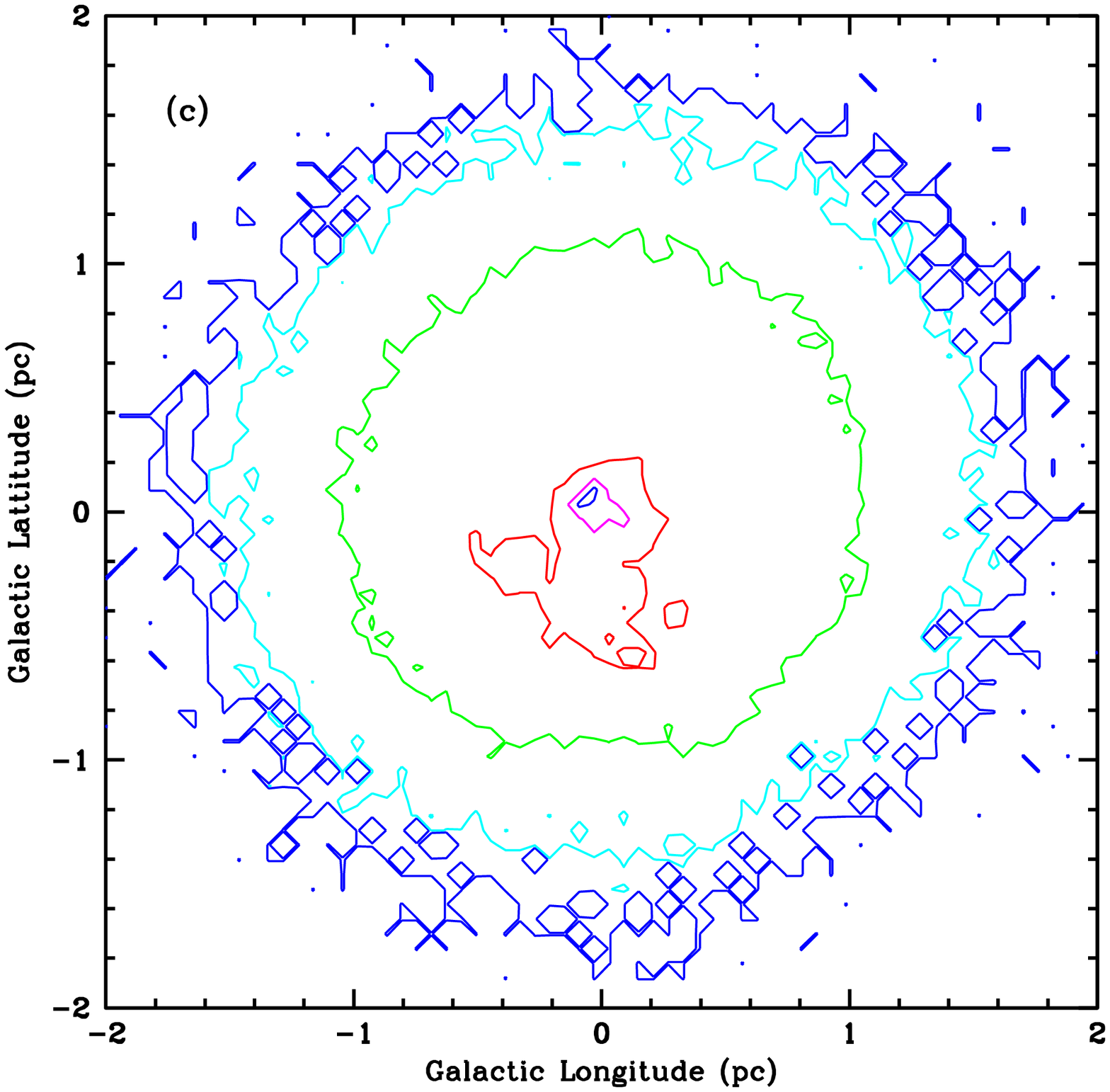}
\epsscale{1.0}
\caption{Same as Fig. 6, but a composite of both our wind profile 
and the inner density set by Rockefeller et al. (2004) using the 
full description of the opacity in equation~\ref{eq:pp}.}
\label{fig:plot7}
\end{figure}
\clearpage

\begin{figure}
\plottwo{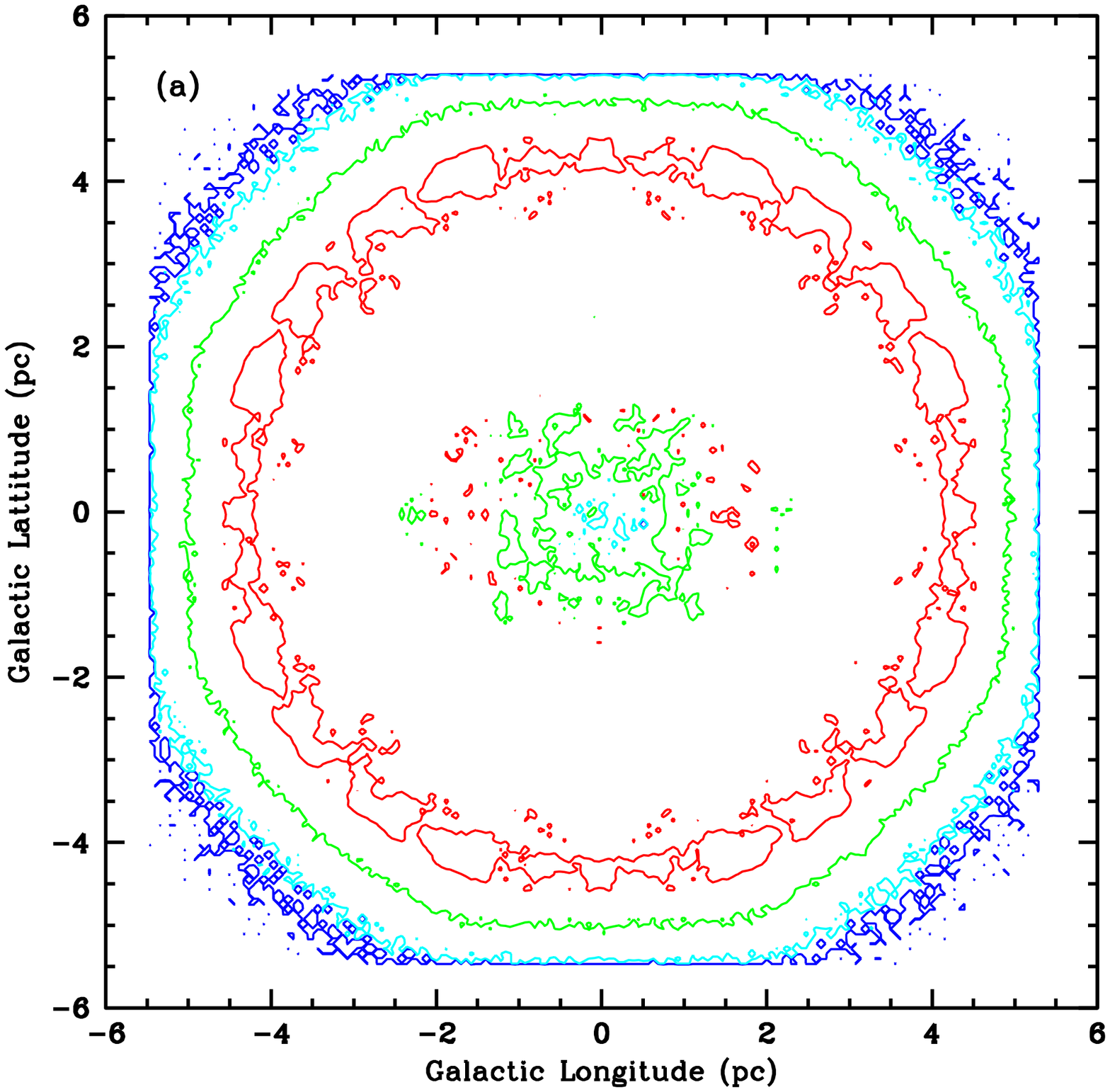}{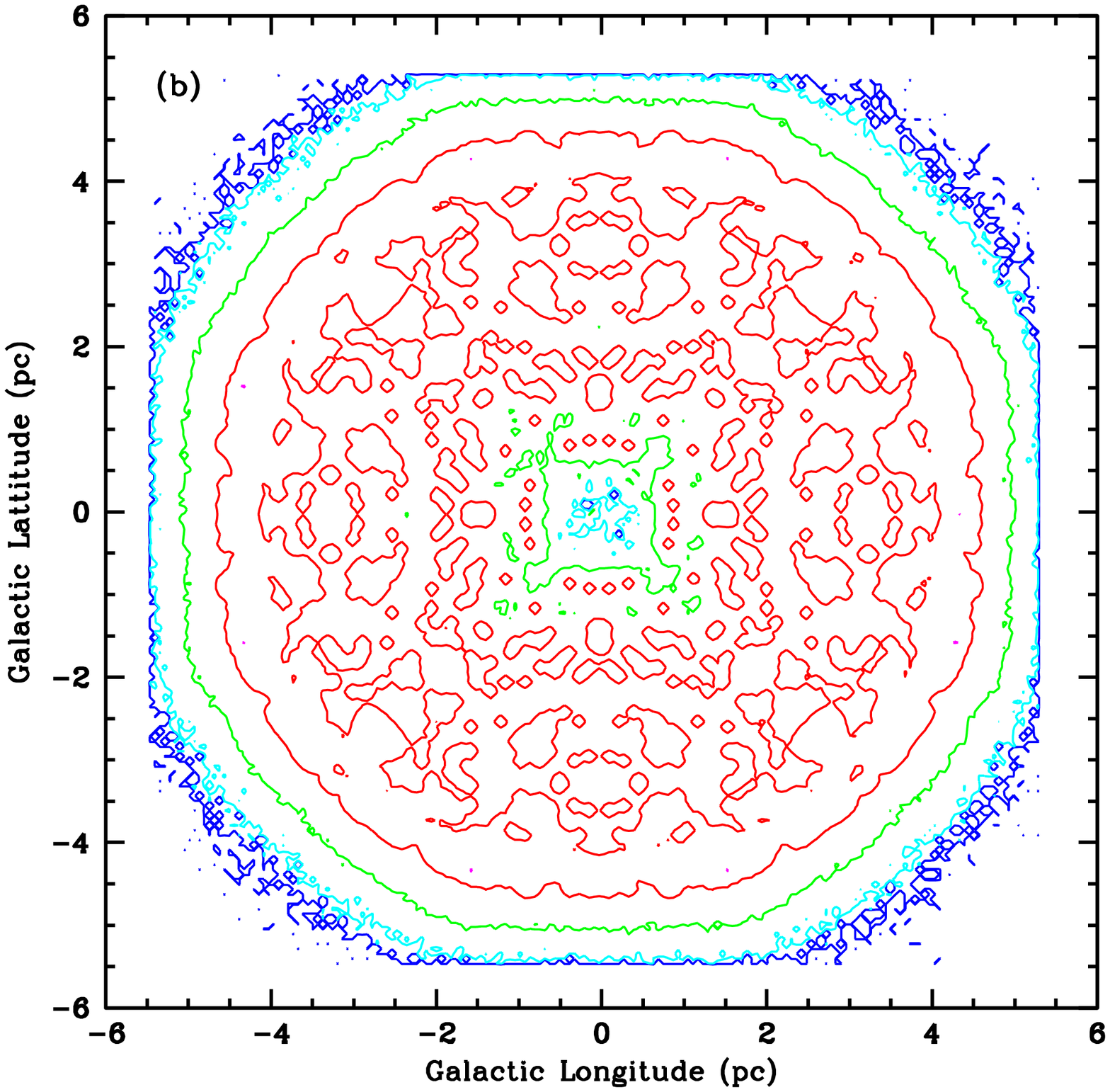}
\epsscale{0.5}\plotone{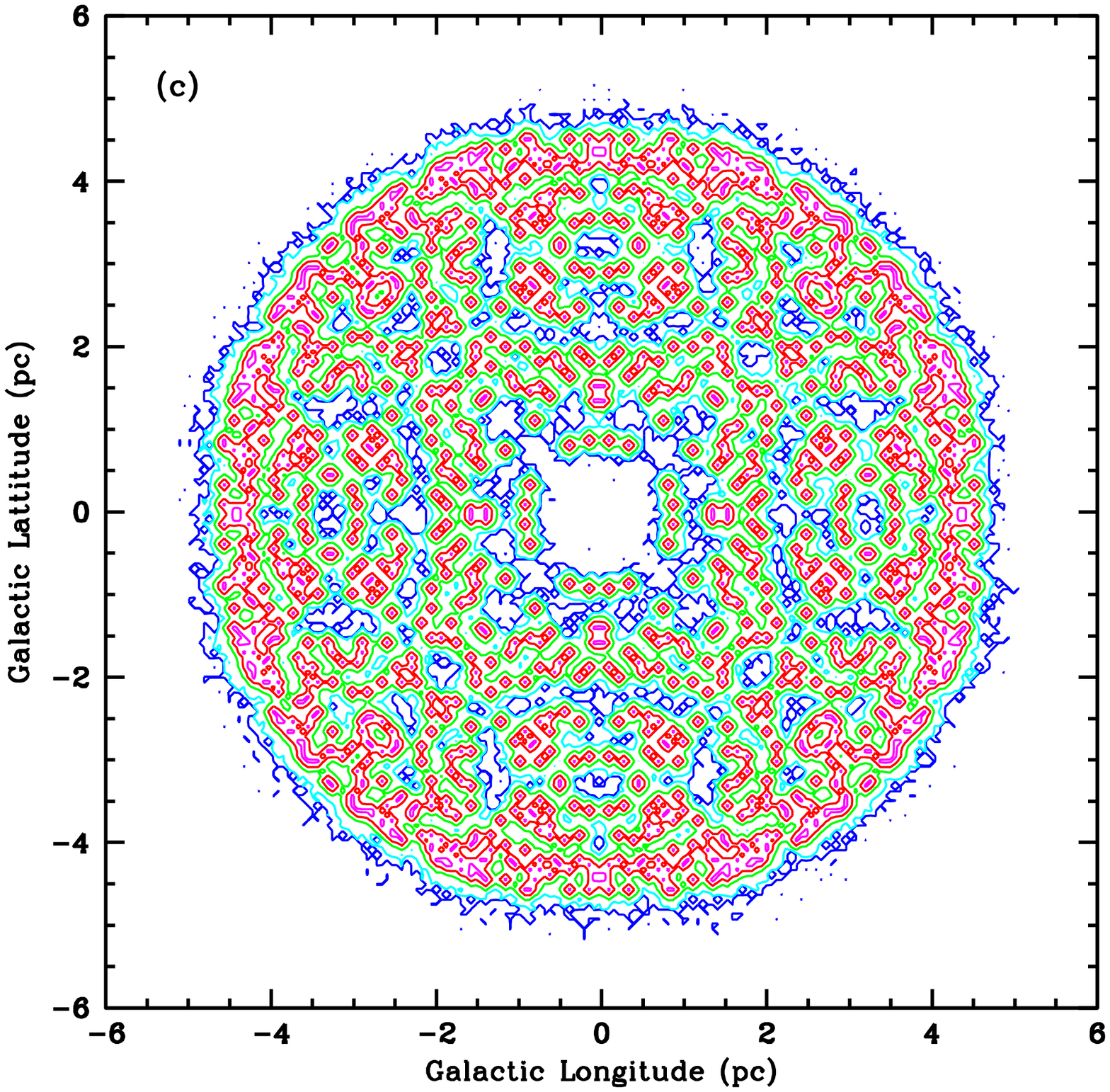}
\epsscale{1.0}
\caption{Same as Fig. 6, but a composite of both our wind profile 
and the inner density set by Rockefeller et al. (2004) using the 
supernova remnant Sgr A East as the source of protons.  Note that 
this simulations produces a much more broad distribution of pion 
production.}
\label{fig:plot8}
\end{figure}
\clearpage

\begin{figure}
\plotone{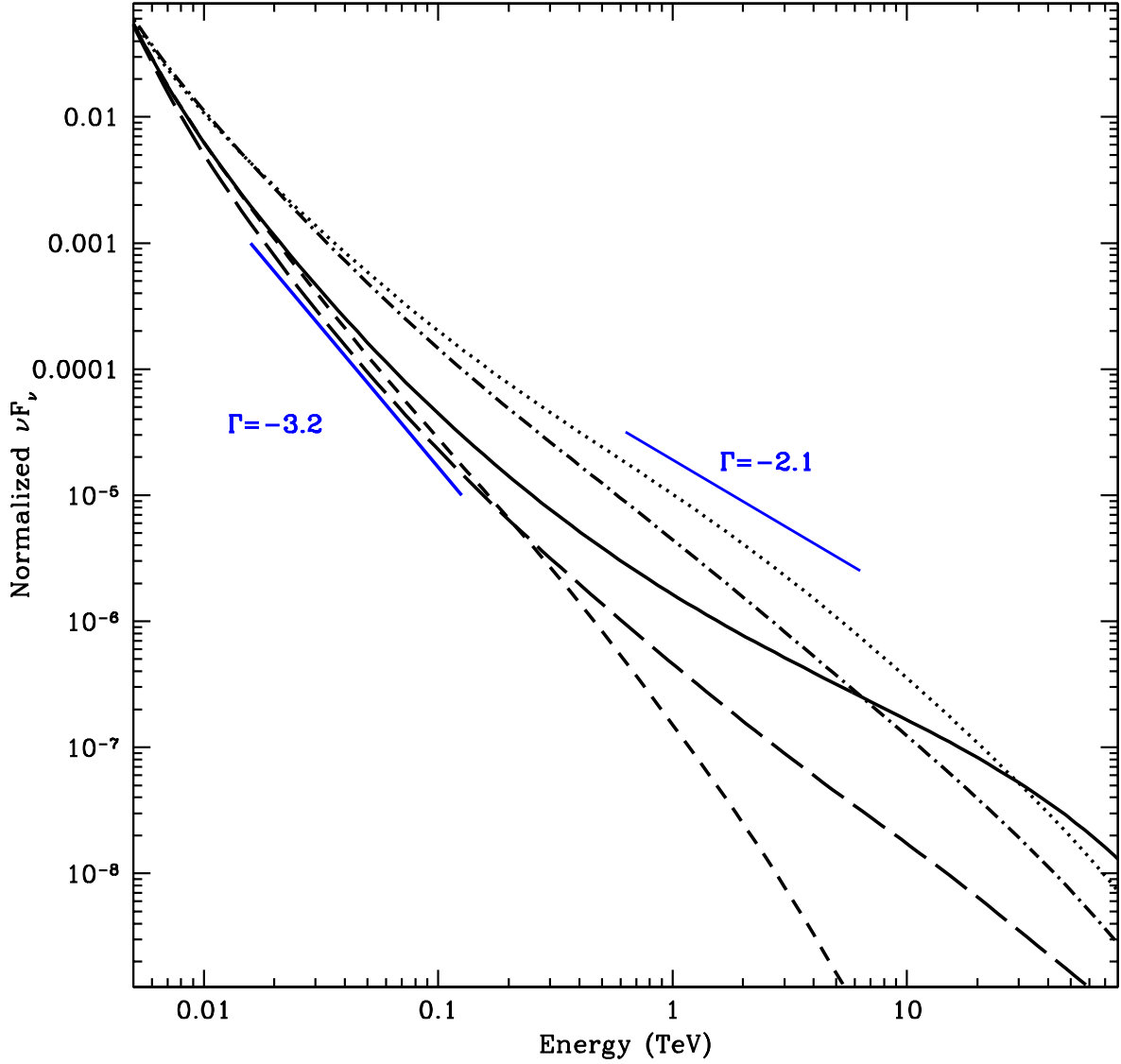}
\caption{Normalized high-energy $\nu F_\nu$ fluxes per unit area as a
function of energy (see Fig. 11) for 5 different models: our standard
model from Fig. 11 (solid), a model only with the wind-blown bubble
component (dotted), and a model with the same density profile, but
with an index for the injection proton energy distribution of -3.3
(dashed), the model using the supernova remnant as a source
(dot-dashed), and the model assuming $q=2$ for the magnetic field
distribution (long-dashed).  Clearly, if the proton energy
distribution does not have a power index near -2.0, we can not get the
correct slope for the gamma-rays.}
\label{fig:specsuite}
\end{figure}
\clearpage

\begin{figure}
\plotone{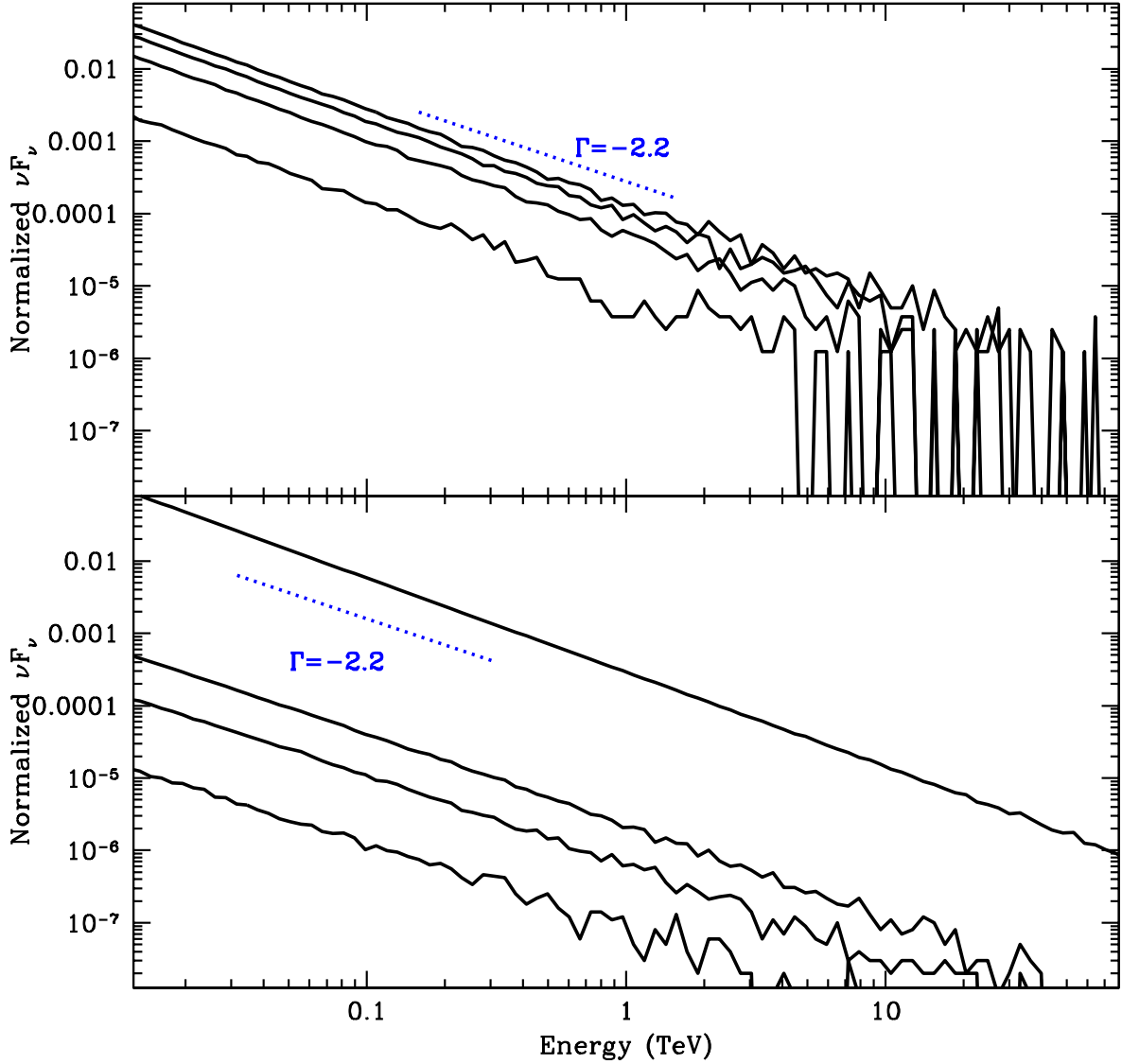}
\caption{Normalized proton $\nu F_\nu$ fluxes per unit area as a
function of energy for a range of separations from the GC ranging from
0.24~pc to 2.4~pc for our standard model (bottom) and the model
assuming $q=2$ for the magnetic field distribution (top).  Although
the index does not change, as we move out, the signal gets
increasingly noisy as the proton scattering rate decreases.}
\label{fig:protspat}
\end{figure}
\clearpage

\begin{figure}
\plotone{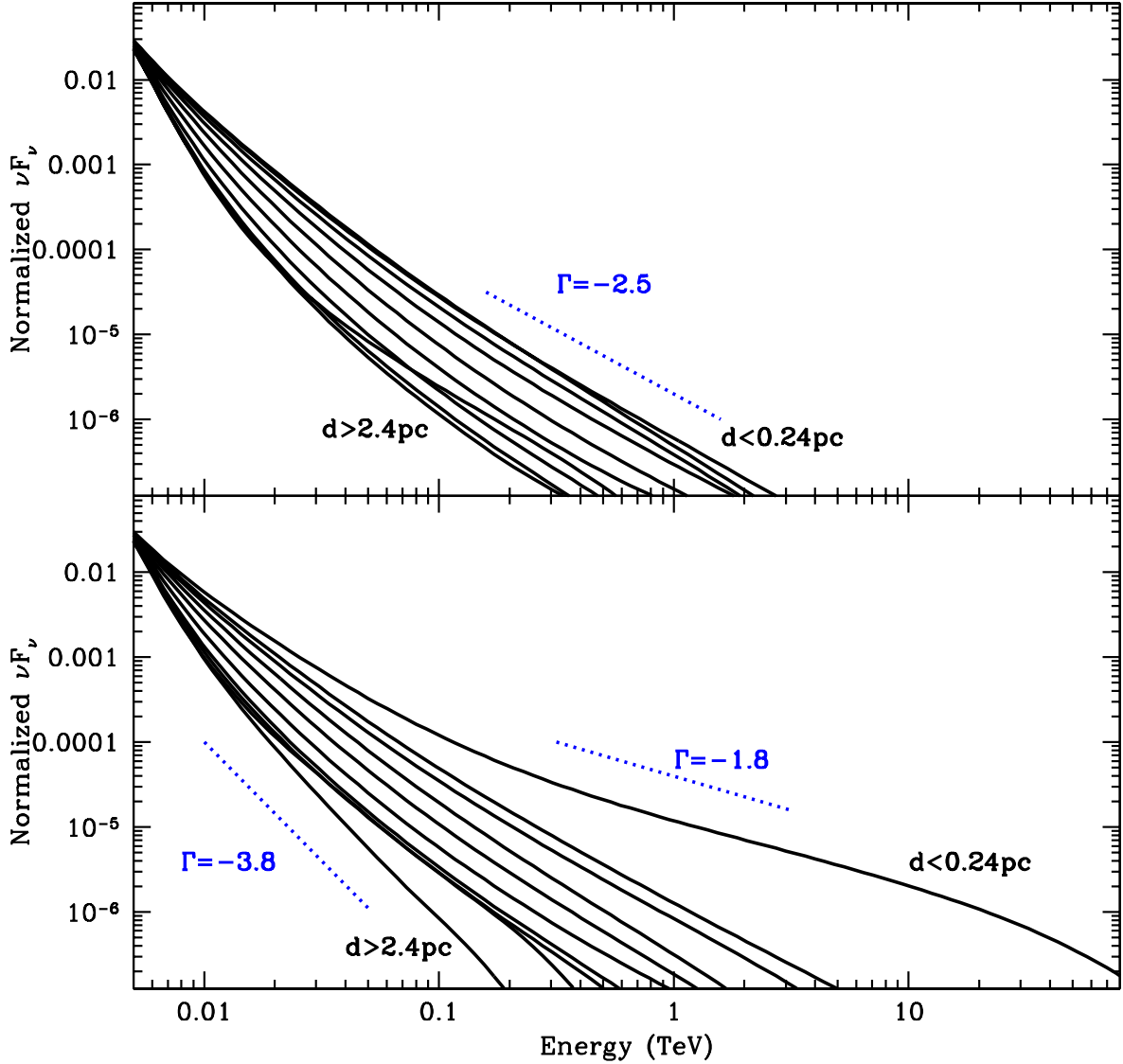}
\caption{Normalized high-energy $\nu F_\nu$ fluxes per unit area as a
function of energy for a range of separations from the GC ranging from
0.24~pc to 2.4~pc for our standard model (bottom) and the model
assuming $q=2$ for the magnetic field distribution (top).  Very close
to the Galactic Center, high energy protons are absorbed, causing a
flattening of the spectrum (this is especially evident in the standard
model).  As we move outward, the density and magnetic field drops
sufficiently that essentially no high-energy protons are absorbed and
the slope of the energy steepens.  However, since we can only observe
the total spectrum at this point, we can not constrain our opacities
with the current observations.}
\label{fig:specspat}
\end{figure}
\clearpage

\begin{figure}
\plotone{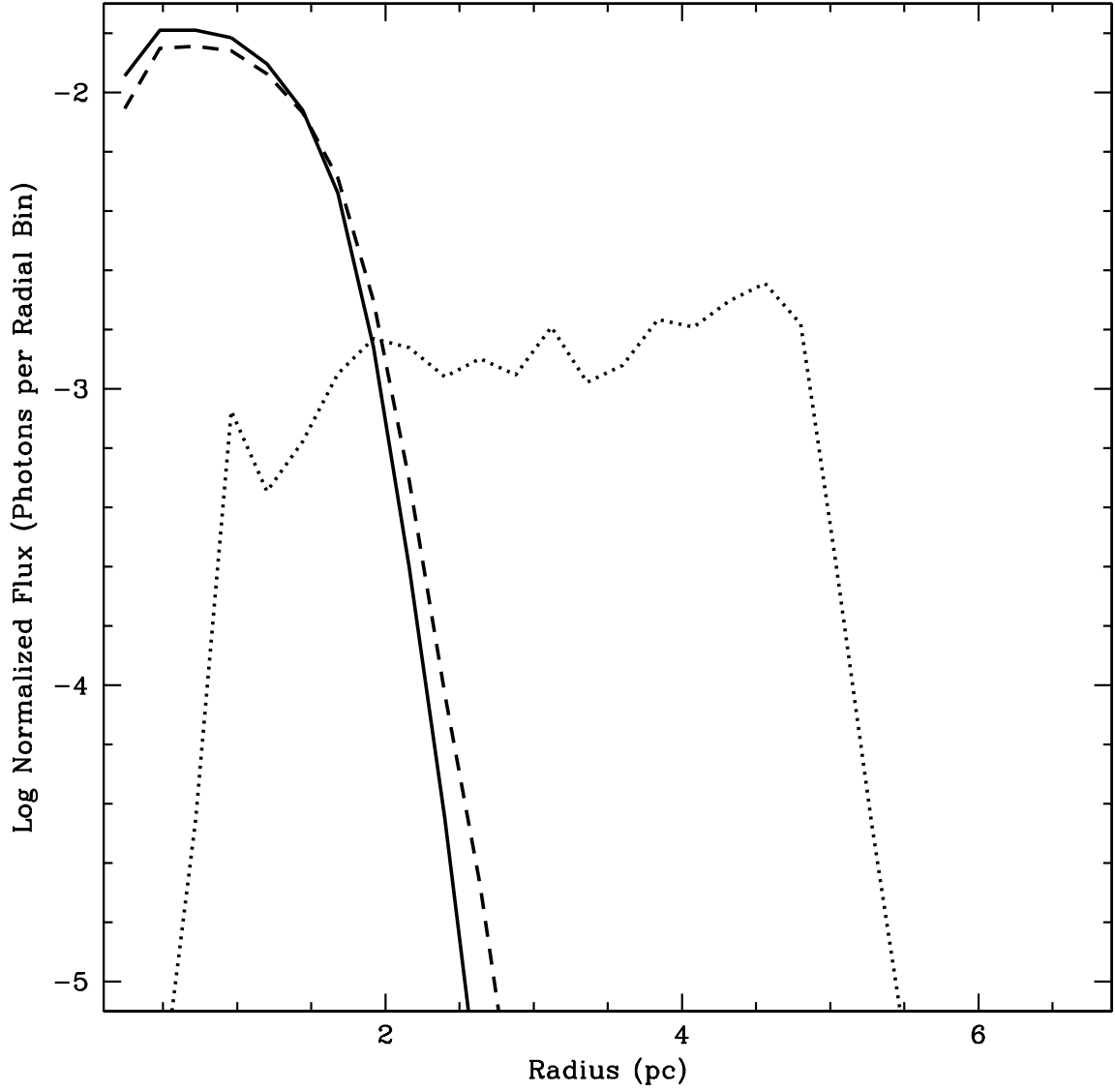}
\caption{Normalized flux per radial bin as a function of radius for
our standard (solid), $q=2$ for the magnetic field distribution
(dashed) and Sgr A East source (dotted) simualtions.  The Sgr A East 
gamma-ray emission is much more extended.}
\end{figure}
\clearpage


\begin{thebibliography}{}

\bibitem[Aharonian et al.(2002)]{Aha02} Aharonian, F.A., Belyanin, A.A., 
Derishev, E.V., Kocharovsky, V.V., \& Kocharovsky, Vl.V. 2002, PRD, 66, 
023005 

\bibitem[Aharonian et~al.(2004)]{Aharonian2004} {Aharonian}, F.~A. 
et~al. 2004, \aap, 425, 13

\bibitem[Aharonian et~al. 2005]{Aharonian2005} {Aharonian}, F.~A. 
and Neronov, A. 2005, \apj, 619, 306

\bibitem[Aharonian et~al. 2006]{Aharonian2006} {Aharonian}, F.~A., et al. 2006, 
Nature, 439, 695

\bibitem[Albert et al. 2006]{Albert06} Albert, J., et al. 2006, ApJ, 638, L101

\bibitem[Atoyan \& Dermer 2004]{AD04} Atoyan, A., \& Dermer, C. D. 2004, ApJ, 617, 
L123

\bibitem[Blasi \& Colafrancesco(1999)]{BC99} Blasi, P., \&
Colafrancesco, S.  1999, Astropart. Phys., 12, 169

\bibitem[Brown, Hartmann, \& Burton(1995)]{BHB95} Brown, A.G.A., 
Hartmann, D., \& Burton, W.B. 1995, A\&A, 300, 903

\bibitem[Chevalier \& Li(1999)]{CL99} Chevalier, R.A., \& Li, Z.-Y. 1999, 
ApJ, 520, L29 

\bibitem[Fatuzzo and Melia(2003)]{Fatuzzo03} Fatuzzo, M. and Melia, 
F. 2003, \apj, 596, 1035

\bibitem[Fryer et al.(1999)]{Fry99} Fryer, C.L., Benz, W., 
Herant, M., \& Colgate, S.A. 1999, ApJ, 516, 892

\bibitem[Fryer, Rockefeller \& Young(2006)]{FRY06} Fryer, C.L.,
Rockefeller, G., \& Young, P.A. 2006, accepted by ApJ

\bibitem[Fryer et al.(2006)]{Fry06b} Fryer, C.L., Rockefeller, G.,
Hungerford, A.L., Melia, F. 2006, ApJ, 638, 786

\bibitem[Heger et al.(2003)]{Heg03} Heger, A., Fryer, C.L., Woosley, S.E., 
Langer, N., \& Hartmann, D.H. 2003, ApJ, 591, 288

\bibitem[Herrnstein \& Ho(2004)]{HH04} Herrnstein, R.M., \& Ho, P.T.P. 
2004, ANS, 1, 583

\bibitem[Herrnstein \& Ho(2005)]{HH05} Herrnstein, R.M., \& Ho, P.T.P. 
2005, ApJ, 620, 287

\bibitem[Hinton et a. 2006]{Hin06} Hinton, J. for H.E.S.S. Collaboration 2006, 
astro-ph/0607351

\bibitem[{Kosack} et~al.(2004)]{Kosack2004} Kosack, K. and Collaboration,
t.~V. 2004, astro-ph/0403422

\bibitem[LaRosa et al.(2005)]{LaR05} LaRosa, T.N., Brogan, C.L.,
Shore, S.N., Lazio, T.J., Kassim, N.E., Nord, M.E. 2005, ApJ, 626, L23

%\bibitem[Liu, Petrosian, \& Melia(2004)]{Liu04} Liu, S., Petrosian, V.,
%and Melia, F. 2004, \apjl, 611, L101

\bibitem[Liu, Melia, \& Petrosian(2006)]{Liu06} Liu, S., Melia, F.,
and Petrosian, V. 2006, \apjl, 636, L798

\bibitem[Mezger et al.(1989)]{Mez89} Mezger, P.G., Zyka, R., Salter, C.J., 
Wink, J.E., Chini, R., Kreysa, E., \& Tuffs, R. 1989, ApJ, 209, 337

\bibitem[Nomoto et al.(2004)]{Nom04} Nomoto, K., Maeda, K., Mazzali,
P.A., Umeda, H., Deng, J., \& Iwamoto, K., in ``Stellar Collapse'',
Astroph. \& Space Science Library, ed. Chris Fryer, Kluwer Academic
Publishers (Dordrecht)

\bibitem[Padoan et al.(2004)]{Pad04} Padoan, P., Jimenez, R., Juvela, M., 
\& Nordlund, A. 2004, ApJ, 604, L49

\bibitem[Petrosian \& Liu 2004]{PL04} Petrosian, V., \& Liu, S. 2004, ApJ, 610, 550

\bibitem[Post 1956]{Po56} Post, R. F. 1956, Rev. Mod. Phys., 28, 338

\bibitem[Priddey et al.(2006)]{Pri06} Priddey, R.S., Tanvir, N.R.,
Levan, A.J., Fruchter, A.S., Kouveliotou, C., Smith, I.A., \& Wijers,
R.A.M.J. 2006, MNRAS, 369, 1189

\bibitem[Quataert \& Loeb 2005]{QL05} Quataert, E., \& Loeb, A. 2005, ApJ, 635, 45

\bibitem[Rockefeller, Fryer, Melia, \& Warren (2004)]{Rock04}
Rockefeller, G., Fryer, C. L., Melia, F., and Warren, M. S.
2004, \apj, 604, 662

\bibitem[Rockefeller et al.(2005)]{Roc05} Rockefeller, G., Fryer, C.L., 
Baganoff, F.K., \& Melia, F.

\bibitem[{Tsuchiya} et~al. 2004]{Tsuchiya2004} {Tsuchiya}, K. 
et~al. 2004, ApJ,  606, L115

\bibitem[Wang et al. 2006]{Wang06} Wang, Q. D., Lu, F. J., \& Gotthelf, E. V. 2006, 
MNRAS, 367, 937 

\bibitem[Yan \& Lazarian 2002]{YL02} Yan, H., \& Lazarian, A. 2002, Phys. Rev. 
Lett., 89, 281102

\end{thebibliography}
\end{document}